\documentclass[aps,pre,twocolumn,amssymb,amsmath,nofootinbib]{revtex4-1}
\usepackage{graphicx,latexsym,pifont,amsmath,dsfont} %
\usepackage{tikz}

\usepackage{times}

\newcommand{\checked}{}

\usetikzlibrary{snakes}
\usepackage{color}
\definecolor{Blue}{rgb}{0.00, 0.00, 1.00}
\definecolor{Red}{rgb}{1.00, 0.00, 0.00}

\newcommand{\rmd}{{\mathrm{d}}}

\newcommand{\half}{\frac12}

\newcommand{\nn}{\nonumber}

\newcommand{\fig}[2]{\includegraphics[width=#1]{#2}}
\newcommand{\Fig}[1]{\includegraphics[width=8.7cm]{#1}}

\newlength{\bilderlength}

\newcommand{\Eqref}[1]{Eq.~\eqref{#1}}

\renewcommand{\epsilon}{\varepsilon}

\def\be{\begin{equation}}
\def\ee{\end{equation}}

\def\bal{\begin{align}}
\def\eal{\end{align}}

\def\bea{\begin{eqnarray}}
\def\eea{\end{eqnarray}}
\renewcommand{\log}{\ln }

\usepackage{color}

\begin{document}

\title{Extreme-Value Statistics of Fractional Brownian Motion Bridges}

\author{Mathieu Delorme and Kay J\"org Wiese}
\address{CNRS-Laboratoire de Physique Th\'eorique de l'Ecole
Normale Sup\'erieure,  
PSL Research University, Sorbonne Universit\'es, UPMC, 24 rue Lhomond, 75005 Paris, France.
}

\begin{abstract}
Fractional Brownian motion   is a self-affine, non-Markovian and translationally invariant  generalization of Brownian motion, depending on the Hurst exponent $H$. 
 Here we investigate   fractional Brownian motion where both the starting and the end point are zero, commonly referred to as bridge processes. Observables are the time $t_+$ the process is positive,   the maximum $m$ it achieves, and the time $t_{\rm max}$ when this maximum is taken. 
 Using a perturbative expansion around Brownian motion ($H=\frac12$), we give the first-order result for the probability distribution of these three variables, and the joint distribution of $m$ and $t_{\rm max}$. Our analytical results are tested, and found in excellent agreement, with extensive numerical simulations, both for $H>\frac12$ and $H<\frac12$.  This precision is achieved by sampling processes with a free endpoint, and then converting each realization to a bridge process, in generalization to what is usually done for Brownian motion. 
 
\end{abstract}

\maketitle

\section{Introduction}
Stochastic processes  are a  powerful tool to describe the evolution of systems where the  microscopic dynamics is not accessible. As an example,   Brownian motion, {\em aka} the  Wiener process, was introduced as an effective probabilistic description for the dynamics of a particle subjected  to  collisions with its environment \cite{Einstein1905}, be it a gas or a liquid.

An important class of such processes, which contain Brownian motion, are 
{\em Markov chains}. For these the  evolution depends only on the current position, but is independent of previous ones. Stated differently: In a discrete dynamics  the increments between successive positions are independent random variables. This {\em Markov property} is at the center of powerful tools \cite{FellerBook} for   stochastic processes, as Green-function methods, the Fokker-Plank equation, etc.

Though Markov chains  successfully   model many systems, there are also  relevant systems which do not evolve with independent increments, and thus are non-Markovian, i.e.\ history dependent. 
Such processes 
naturally appear for  the  dynamics of a single point in a spatially extended object, as for instance    a single spin in a magnet, or   a marked monomer in a polymer. Their dynamics becomes non-Markovian due to the coupling to the neighbors. 

Dropping the Markov property, but keeping the other ingredients of Brownian motion, i.e.\ Gaussianity, scale invariance and stationarity of the increments defines an enlarged class of random processes, known as fractional Brownian motion (fBm), and parameterised by the {\em Hurst parameter} $H$, which  quantifies the self-affinity of the process. Its covariance function is
\begin{equation}\label{covariance}
G^{-1}(t_1,t_2)=\langle X_{t_1} X_{t_2} \rangle = t_1^{2H} + t_2^{2H} - |t_1-t_2|^{2H} \ .
\end{equation}
Since the process is Gaussian, \Eqref{covariance} defines  it. 
 Such processes appear in a broad range of contexts: Anomalous diffusion \cite{BouchaudGeorges1990}, diffusion of a marked monomer inside a polymer 
\cite{WalterFerrantiniCarlonVanderzande2012,AmitaiKantorKardar2010}, 
polymer translocation through a pore \cite{AmitaiKantorKardar2010,ZoiaRossoMajumdar2009,DubbeldamRostiashvili2011,PalyulinAlaNissilaMetzler2014}, single-file diffusion 
\cite{KrapivskyMallickSadhu2014,KrapivskyMallickSadhu2015,KrapivskyMallickSadhu2015a} observable experimentally  in ion channels \cite{KuklaKornatowskiDemuthGirnusal1996,WeiBechingerLeiderer2000}, the dynamics of a tagged monomer \cite{GuptaRossoTexier2013,Panja2011}, finance (fractional Black-Scholes, fractional stochastic volatility models, and their limitations) \cite{CutlandKoppWillinger1995,Rogersothers1997,RostekSchobel2013}, hydrology  \cite{MandelbrotWallis1968,MolzLiuSzulga1997},  and many more. 
Their extreme-value statistics has been studied in many referenes \cite{Sinai1997,Majumdar1999,Molchan1999,NourdinBook,Aurzada2011,OerdingCornellBray1997,KrapivskyMallickSadhu2014,KrapivskyMallickSadhu2015,KrapivskyMallickSadhu2015a}.

\begin{figure}[t]\Fig{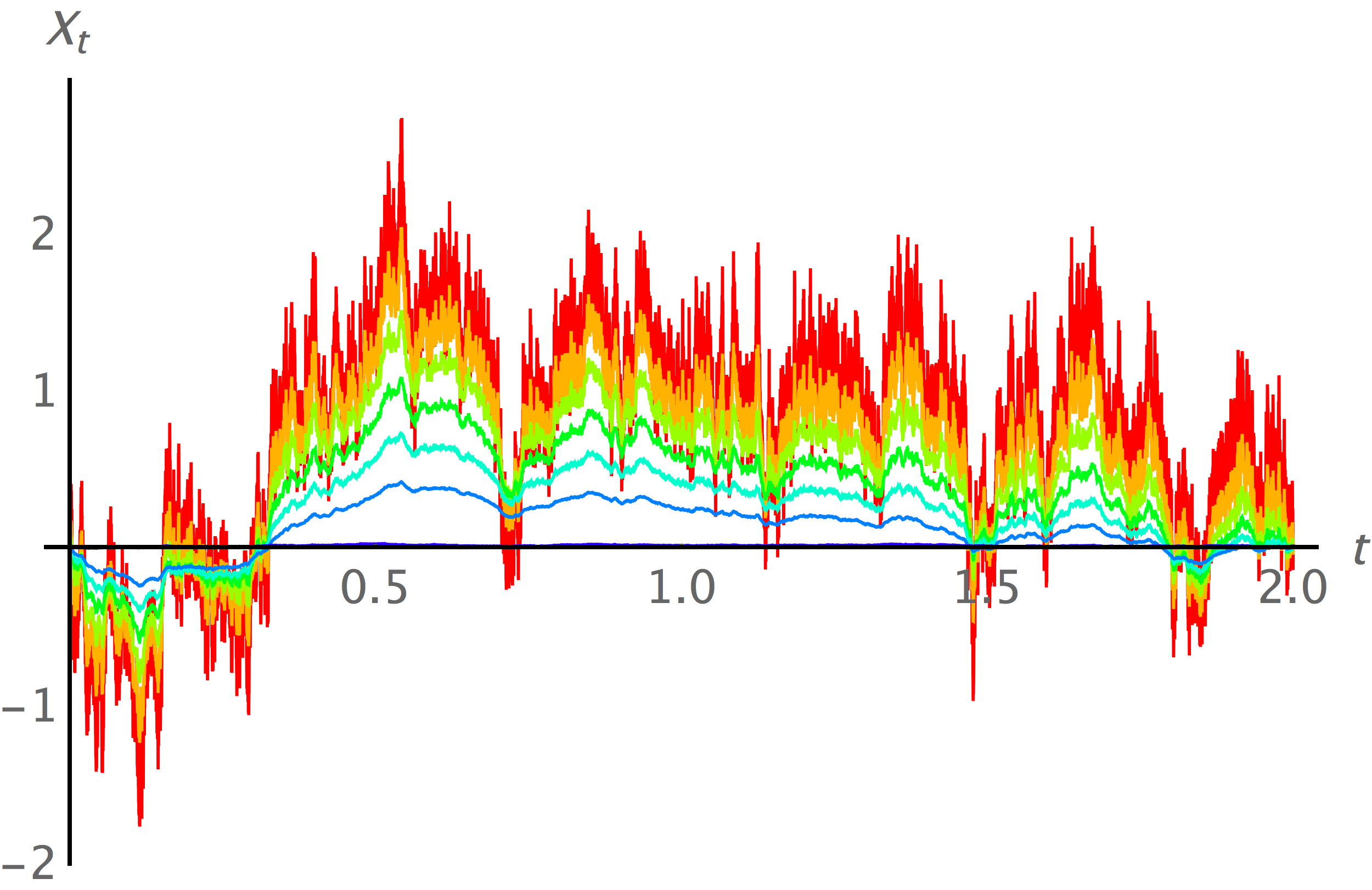}
        \caption{Exemples of fBm bridges for different values of $H$, generated from the same random numbers using the Davis and Harte procedure \cite{DiekerPhD}. $H=0.25$ in red (outmost curves)  to $H=0.875$ in blue (innermost), with increments of $1/8$.}
\label{f:differentHs-bridge}
\end{figure}

When studying random processes in a time interval $[0,T]$, quite generally the initial value $X_0$ is known, and the endpoint $X_T$ is itself a random variable determined by the random process. On the other hand, there are also cases when one  knows the endpoint $X_T$. These processes are   referred to as {\em bridge processes} or  {\em bridges}. For a Brownian one refers to {\em Brownian bridges}.

Using a Fourier decomposition with the same amplitude for each mode, but different values of $H$, one can generate realizations of fBm bridges, and study their dependence on $H$, see Ref.~\cite{DiekerPhD} and section \ref{s:GaussianBridge} below.  
Sample trajectories ranging from $H=0.25$ (red) to $H=0.875$ (blue) in increments of $0.125$ are presented on Fig.~\ref{f:differentHs-bridge}.

   Bridges are useful building blocks in constructing more complicated observables; we will see an application of this idea in section \ref{s:PositiveTime} below. 
They are also commonly used 
in constructing refinements of random walks, {e.g.}\  for financial modeling  \cite{AndersenPiterbargBook1}. Finally, they appear as the difference  from the asymptotic limit in the construction of the {\em empirical distribution function} \cite{Van-der-VaartBook}. 

\medskip

We investigate three observables relevant for bridges: \begin{enumerate}
\item[(i)] the time $t_{\rm max}$ the random process achieves its maximum,
\item[(ii)]  the value $m$ of this maximum,
\item[(iii)] the time $t_+$ the process is positive,
aka its {\em positive time}, supposing one starts at $X_0=0$. 
\end{enumerate}
For Brownian motion, and 
for a bridge terminating at its starting point after time $T$, both $t_{\rm max}$ and $t_{+}$ have a uniform distribution  \cite{Levy1940}
\be \label{BrownianBridgeArcsinLaws}
{\cal P}_{H=1/2}^{\rm bridge}(t_{\rm max})  = {\cal P}_{H=1/2}^{\rm bridge}(t_{+}) = \frac{1}T\ . 
\ee
In contrast, for Brownian motion with a free endpoint (i.e.\ without constraint)  the corresponding  probability reads \cite{FellerBook,Levy1940}
\be \label{BrownianArcsinLaws}
{\cal P}_{H=1/2}^{\rm free}(t=t_{\rm max})  = {\cal P}_{H=1/2}^{\rm free}(t=t_{+}) = \frac{1}{\pi \sqrt{t(T-t)}}   \ .
\ee
 These two results, as well as a way to interpolate between them for  the positive-time distribution  can be found in Ref.~\cite{NikitinOrsingher2000}. For the maximum value $m$, up to time $T$, the probability distributions are
\begin{align}
{\cal P}_{H=1/2}^{\rm bridge}(m)   &= \frac{2 m}{T}e^{-\frac{m^2}{T}} \Theta(m)\ , \label{MaxDistribBridge}\\
{\cal P}_{H=1/2}^{\rm free}(m) &= \frac{e^{-\frac{m^2}{4T}}}{\sqrt{\pi T}}  \Theta(m)\ .
\end{align}
Properties of    fractional Brownnian motion were recently investigated within a perturbative approach in  $H=1/2+\varepsilon$, expanding  around   Brownian motion, corresponding to $H=1/2$ \cite{WieseMajumdarRosso2010,DelormeWiese2015,DelormeWiese2016}. We extend these results   by considering  {\em bridge processes}. 
While observables related to the maximum of  an unconstrained fractional Brownian motion were already considered in  Refs.~\cite{DelormeWiese2015,DelormeWiese2016}, the observable   $t_+$ is considered for the first time here.
Indeed, we will show that at leading order in $\epsilon=H-\frac12$, the probability distributions for $t_{\rm max}$ and $t_+$ are different, contrary to  Brwonian motion, and  processes with a free endpoint, where they agree at leading order \cite{DelormeWieseUnpublished}.

Finally we   test our analytical results against numerical simulations for $H=0.4$, $H=0.6$, and $H=0.66$. This is achieved by constructing a {\em subtracted} process out of each realization of a fBm with free endpoints. This procedure yields the same statistics as a fractional Brownian bridge, and    is much more efficiently simulated than   an unconstrained fBm, for which one retains only realizations which are bridges.  

\medskip
%

This article is organised as follows: Section \ref{s:GaussianBridge} introduces some general results about Gaussian bridges, as well as their application to fractional Brownian motion.

Section \ref{s:Perturbative} recalls   the methodology developed in Ref.~\cite{DelormeWiese2016} on the perturbative expansion around  Brownian motion.

Section \ref{s:PositiveTime} introduces $t_+$, the time spent    by the process in the positive half space. We start with  a discrete random walk before taking the continuum   limit to obtain the distribution of $t_+$ for  Brownian motion. This is used as a starting point for the perturbative expansion described in the previous section, with some technical steps left to appendix \ref{a:Abel}. The analytical results obtained are then compared to numerical simulations.

Section \ref{s:ExtremeValues} presents   results on  the extreme-value statistics for a fBm bridge: the maximum value $m$ as well as the time $t_{\rm max}$ to reach it. Some of these results are derived from a general calculation performed in Ref.~\cite{DelormeWiese2016}; we also  present a new and   simpler way to obtain the maximum-value distribution.

Several appendices complete our work: Appendix \ref{a:Abel} contains details about the inverse of an integral transform appearing in our calculation, and its relation to the Abel transform.

Appendix \ref{a:inv-Lap-trafos} summarises  the necessary inverse Laplace transforms needed in the main text.

\section{Preliminaries: Gaussian Bridges}
\label{s:GaussianBridge}

Consider a   real-valued process $X_t$, starting at $X_0=0$. We define a {\em  bridge}, denoted   $X_t^{\rm B}$, to be the same process {\em conditioned} to be at $a$ at time $T$.  Its one- and two-point correlation functions are
\begin{align}
\langle X_{t_1}^{\rm B} \rangle &= \frac{\langle X_{t_1}  \delta(X_T-a)\rangle}{\langle \delta(X_T-a)\rangle}\ ,
        \label{Bridge_def1} \\
        \langle X_{t_1}^{\rm B} X_{t_2}^{\rm B} \rangle &= \frac{\langle X_{t_1} X_{t_2} \delta(X_T-a)\rangle}{\langle \delta(X_T-a)\rangle}\ .
        \label{Bridge_def}
\end{align}
We now assume  that $X_t$ is a centered Gaussian process, i.e.\ $\langle X_t \rangle =0$ for all $t$, and that cumulants of order higher than $2$ vanish. To express the correlation function of the bridge process in terms of the unconditioned process, we insert the identity $ \delta(x)  = \int_{-\infty}^{\infty} e^{i k x} \frac{\rmd k}{2\pi}$ into the above equations. After some lines of algebra    presented in appendix \ref{a:bridge-details}, we arrive at
\begin{align} \label{10}
 \langle X_{t_1}^{\rm B}   \rangle &= a \frac{\langle X_{t_1} X_{T}\rangle}{\langle X_T^2\rangle} \\
\label{11}  \langle X_{t_1}^{\rm B} X_{t_2}^{\rm B} \rangle &= \langle X_{t_1} X_{t_2}\rangle - \Big[ \langle X_{T}^2\rangle-a^2  \Big] \frac{\langle X_{t_1} X_{T} \rangle\langle X_{t_2} X_{T} \rangle}{\langle X_{T}^2\rangle^2}\ .
\end{align} 
Consider now the {\em subtracted} process $X_t^{\rm S}$ defined from the original process $X_t$ as
\begin{equation}\label{subtractedProcess}
X_t^{\rm S}:=X_t - (X_T-a) \frac{\langle X_{t} X_{T} \rangle}{\langle X_{T}^2 \rangle}\ .
\end{equation}
One easily checks that  its one and two-point correlation functions    coincide with those of $X_t^{\rm B}$ given in Eqs.~(\ref{10})--(\ref{11}). 
This is sufficient to conclude that $X_t^{\rm B}$ and $X_t^{\rm S}$ are the same processes, 
\begin{equation}\label{SubsBridgeEquivalence}
X_t^{\rm S}  \overset{\rm law}{=} X_t^{\rm B}\ .
\end{equation}
While this result was   derived in Ref.~\cite{GasbarraSottinenValkeila2005} by other methods, the prescription (\ref{subtractedProcess}) does not seem to be generally known.

Frequently used for Brownian motion $X_t:=B_t$ 
 the subtracted process \eqref{subtractedProcess} reduces to
\begin{equation}
B_t^{\rm S}=B_t- \frac{t}{T} \Big( B_T -a \Big)\ .
\end{equation}
This
 is  equivalent in law to a Brownian bridge ending at $a$. 

For  fractional Bronwian motion with Hurst exponent $H$, the subtracted term is non-linear in $t$,    containing the   expression 
\begin{equation}
f\!\left(\vartheta:=\frac{t}{T}\right):= \frac{\langle X_{t} X_{T} \rangle}{\langle X_{T}^2 \rangle} =\frac12\Big[  1 + \vartheta^{2H}-\left(1-\vartheta\right)^{2H}\Big]\ .
\end{equation}
The     equivalence \eqref{SubsBridgeEquivalence} is crucial for the numerical simulations presented in this work. Simulating bridge process using its   definition  requires to discard almost all generated paths, while the subtracted process can be constructed from every generated path without loss of statistics.

\section{Pertubative approach to fBm}
\label{s:Perturbative}

We recall here some useful definitions for   fBm, as well as the ideas of the perturbative expansion around   Brownian motion developed in Refs.~\cite{WieseMajumdarRosso2010,DelormeWiese2016}.

First, as   fractional Brownian motion is a Gaussian process, it is characterized by its covariance function $G^{{-1}}$ given in \Eqref{covariance}. This covariance function  allows us to write   an action  for the possible realizations of $X_t$,
\begin{equation}
S[X]= \frac12 \int_{t_1,t_2} X_{t_1} G(t_1,t_2) X_{t_2}\ .
\end{equation}
This yields the average of any observable $O[X]$ for the fBm within a path-integral formulation,
\begin{equation}
\left\langle O[X] \right\rangle = \int \mathcal{D}[X] \,O[X]\, e^{-S[X]}\ .
\end{equation}
To   compute observables explicitly from this   expression, we   expand the action around   Brownian motion, corresponding to $H=1/2$ in \Eqref{covariance}. Writing $H= 1/2 + \varepsilon$,  we obtain  at first order in $\varepsilon$
\begin{align}\label{ActionExpansion}
S\left[ X \right] =&\;  \int_0^T \rmd t_1\,\frac{\dot{X}_{t_1}^2}{4 D_{\varepsilon,\tau}} \\
& -\frac{\varepsilon}{2} \int_0^{T-\tau} \rmd t_1 \int_{t_1+\tau}^T \rmd t_2 \,\frac{ \dot{X}_{t_1} \dot{X}_{t_2}}{|t_2-t_1|}  +\mathcal{O}(\varepsilon^2)\nn \ .
\end{align}
The first term is the standard action of   Brownian motion, with a rescaled diffusive constant
\begin{equation}\label{DiffusConstant}
 D_{\varepsilon,\tau} = 1+ 2\varepsilon[1+\ln(\tau)] +O(\varepsilon^2)\ .
\end{equation}
The regularisation cut-off $\tau$ (wich is an UV cut-off in time) appears   in the second term of the action, which is a non-local (in time) interaction between derivatives of the process. For the derivation of this expansion  we refer to  Ref.~\cite{WieseMajumdarRosso2010}.

Note  that the non-locality in time of the action is a   manifestation of the non-Markovian nature of fractional Brownian motion.
We will use this formalism to compute observables for bridges of fBm in an $\varepsilon$ expansion, following the   strategy and using results  of Ref.~\cite{DelormeWiese2016}.

\section{Positive time of a fBm Bridge}\label{s:PositiveTime}
In this section, we investigate the distribution of the time spend up to time $T$ by the process $X_t$ in the positive half space. This time, denoted $t_+$, is defined by
\begin{equation}
t_+:=\int_0^T \rmd t\, \Theta(X_t)\ ,
\end{equation}
where $\Theta$ is the Heavyside function, $\Theta(x)=1$ if $x >0$, and $\Theta(x)=0$ otherwise, and $X_0=0$. 

Below, we first consider a discrete random walk   and derive the Laplace transform  (i.e.\ generating function) of the distribution of $t_+$. Taking the continuous-time limit allows us to obtain the distribution of $t_{+}$ for Brownian motion. We use this result to construct our perturbative expansion for a fractional Brownian motion bridge,  and to derive an analytical prediction at order $\varepsilon$.

\subsection{Positive time of a discrete random walk}
Consider a discrete random walk $X_n$ with discrete steps $\pm 1$ (without bias), starting at $X_0=0$.
We denote $N_{n,x}$ the number of paths which goes from $0$ to $x$ in $n$   steps. This number is non-zero only if $x$ and $n$ have the same parity and $x$ is smaller than $n$. 
It can be obtained by retaining the term of order $q^x$ from the  generating function for all paths, $(q+q^{-1})^n$, i.e.\ 
\be
\left(q+\frac1q \right)^{\!n} = \sum_{i=0}^n q^i \left(\frac1q\right)^{n-i} \left({ n \atop i} \right)\ .
\ee
Identifying $x =  2i-n$ yields 
\begin{equation}\label{23}
N_{n,x} =  \left( {n \atop\frac{n+x}{2}}\right)\ .
\end{equation}
It can also be deduced as follows: 
Paths ending in $x$ have $n_+=\frac{n+x}2$ up segments, and $n_- =\frac{n-x}2$ down segments. The number of paths with $n_+$ up segments  is $( {n \atop n_+})  $, which again yields Eq.~(\ref{23}).

\begin{figure}
\Fig{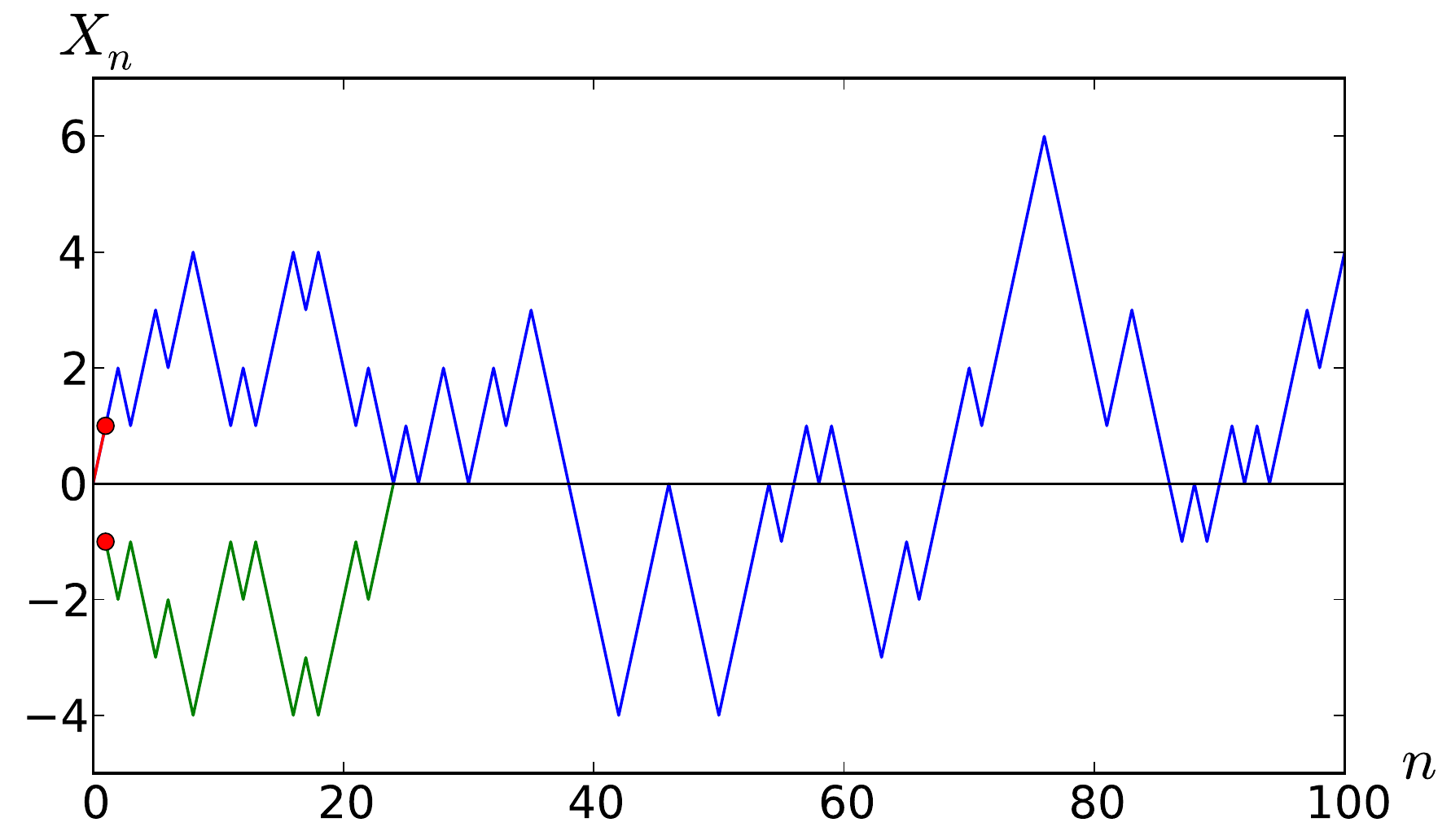}
\caption{Illustration of the reflection principle: Every path emanating from $1$ and attaining zero again (blue) is compensated by a {\em reflected} path emanating from $-1$ (green).}
\label{f:reflection-principle}
\end{figure}

Denote by $N_{n,x}^+$ the number of {\em strictly positive paths}, i.e.~$X_i>0$ for all $i>0$,  which go from $0$ to $x>0$ in $n$ steps. 
By the reflexion principle, illustrated on figure \ref{f:reflection-principle}, this  is the same as the number of paths that go from $1$ to $x$ in $n-1$   steps,  minus the number of paths which start at $-1$ and go to $x$ in $n-1$ steps, 
\begin{equation}
N_{n,x}^+=N_{n-1,x-1}-N_{n-1,x+1}=\frac{x}{n}N_{n,x}\ .
\end{equation}
The ratio
\be
\frac {N_{n,x}^+}{N_{n,x}} = \frac x n
\ee
is the probability that a path from $0$ to $x$ in $n$ steps is {\em strictly positive}, also known as  the Ballot theorem\footnote{The ballot theorem states that if in an election candidate $A$ receives $p$ votes and candidate $B$ receives $q$ votes with $p>q$, the probability that $A$ stays ahead of $B$ throughout the count is $(p-q)/(p+q)$, see Refs.~\cite{Bertrand1887,FellerBook}.}.

Another quantity of interest is the number of {\em excursions}, i.e. paths that go from $X_0=0$ to $X_{2n}=0$  with all intermediate positions   positive, and which we denote $N^{+,\rm first}_{2n}$, because the end point is   the first zero of the path. Such a path necessarily has $X_{2n-1}=1$, which   gives
\begin{equation}\label{PositiveBridgeDiscret}
N^{+,\rm first}_{2n}=N^+_{2n-1,1}=\frac{1}{2n-1}\left(\! {2n-1 \atop n}\!\right)=\frac{(2n-2)!}{n!(n-1)!}\ .
\end{equation}
We now   study the time when a random process is positive: A segment $S_i$ from $i-1$ to $i$ is considered {\em positive} if $X_{i-1}+X_{i} > 0$, and {\em negative} otherwise. 
Note that contrary to the positions $X_i$, a segment $S_i$ is either positive or negative. 
The time $t_+$ a random walk is positive is defined as the number of positive segments. 

Denote $N^{\rm pos}_{2n,2k}$ the number of bridge paths of length $2n$ with $2k$ positive intervals; by convention we set 
 $N^{\rm pos}_{0,0}:=1$.
We can use Eq.~\eqref{PositiveBridgeDiscret} to get a recursion relation for $N^{\rm pos}_{2n,2k}$, with $n\geq 1$,
\begin{equation}\label{Recursion}
N^{\rm pos}_{2n,2k}=\sum\limits_{i=1}^{n} \!\left[ N^{+,\rm first}_{2i}N^{\rm pos}_{2(n{-}i),2(k{-}i)}+N^{+,\rm first}_{2i}N^{\rm pos}_{2(n{-}i),2k}\right]\!.
\end{equation}
This is illustrated on figure \ref{f:reflection-principle}.
In this sum, $2i$ is the position of the first zero (after the origin) of the path of lenght $2n$. Since the path does not change sign these $2i$ first segments are either all positive (first term inside the sum) or   negative (second term). 

To solve this equation, we introduce two generating functions: 
\begin{align}
\tilde p^{\rm pos}(\nu,\rho)&:=\sum\limits_{n \geq 0} \sum\limits_{k \geq 0} \nu^{2k} \rho^{2n} \frac{N^{\rm pos}_{2n,2k}}{2^{2n}}\ ,\\
\label{pfirstGenerating}
\tilde p^{+,\rm first}(\rho)&:= \sum\limits_{n > 0} \rho^{2n} \frac{N^{+,\rm first}_{2n}}{2^{2n}}= \frac{1-\sqrt{1-\rho^2}}{2}\ .
\end{align}
Inserting these definitions  into  \Eqref{Recursion} transforms the recursion relation into an algebraic equation
\begin{equation}\label{30}
\tilde p^{\rm pos}(\nu ,\rho) = \left[\tilde p ^{+,\rm first}(\nu \rho) + \tilde p ^{+,\rm first}( \rho) \right]\tilde p^{\rm pos}(\nu  ,\rho)  +1\ .
\end{equation}
Eq.~(\ref{30}) can be solved as
\be
\tilde p^{\rm pos}(\nu ,\rho)  = \frac1{1-\tilde p ^{+,\rm first}(\nu \rho) - \tilde p ^{+,\rm first}( \rho)}\ .
\ee
This is a geometric sum of the form
\be
\tilde p^{\rm pos}(\nu ,\rho)  = \sum_{n\ge 0} \left[ \tilde p ^{+,\rm first}(\nu \rho) + \tilde p ^{+,\rm first}( \rho)\right]^n\ .
\ee
Its interpretation is simple: All bridges can be constructed as a sequence of first-return bridges. In a first-return bridge each factor of $\rho$ comes with a factor of $\nu$ for the positive paths, and alone for negative paths. 

Using the explicit expression of \Eqref{pfirstGenerating}, we obtain
\begin{equation}\label{PostimeBridgeGenerating}
\tilde p^{\rm pos}(\nu ,\rho)= \frac{2}{\sqrt{1-(\nu \rho)^2}+\sqrt{1-\rho^2}}\ .
\end{equation}
Other   generating functions can be obtained as well:
First,   for the probability to return to zero (including the term with zero steps) the latter is  
\begin{equation} 
\tilde{p}_0(\rho):= \sum\limits_{n\geq 0} \rho^n \frac{N_{n,x}}{2^n}= \frac1{\sqrt{1-\rho^2}}\ .
\end{equation}
For the probability to return to 0 without having become negative, this is (including the term with zero steps)
\bea\label{32}
\tilde{p}_0^{\ge0}(\rho)  &=& \frac1{1-\tilde p ^+_{\rm first}(\rho)} \equiv \tilde p ^{\rm pos}(0,\rho) \nn\\
&=& \frac2{1+\sqrt{1-\rho^2}} \ .
\eea
The   generating function for paths starting at zero and  ending in $x$ without ever returning to zero can be obtained as well
\begin{eqnarray} 
\tilde{p}^+_x(\rho) &:=& \sum\limits_{n\geq 0} \rho^n \frac{N^+_{n,x}}{2^n} \nn\\
&=& \frac{\rho^x}{\left(1+\sqrt{1-\rho^2}\right)^x} =
\frac{\left(1-\sqrt{1-\rho^2}\right)^x}{ \rho^x}\ . ~~~~~~~
\label{36}
\end{eqnarray}
This   can be understood by considering the path from the end: One can first go up and down to the starting value $x$ for a number $n\ge 0$ steps, before going down by one step, leading to $\tilde{p}_0^{\ge0}(\rho)  \times \frac\rho2 $ for the generating function to (backwards!) reach $x-1$. Repeating this $x$ times, and using Eq.~(\ref{32}),  we arrive at  Eq.~(\ref{36}).

\subsection{Propagators in continuous time}

We now wish to take the continuum limit. To this aim, we note that in the limit of a time-discretisation step ${\delta t\to 0}$, the process 
\be\label{RandomWalkLimit}
X_t \simeq  \sqrt{2\delta t} X_{n}\ , \mbox{~~ with~~ } n = \mbox{floor}\left( \frac t{\delta t} \right)
\ee
converges to a Brownian. 
The normalisation ensure that  we recover the covariance function \eqref{covariance} with $H=\frac12$. 

Denote by ${\cal P}(t_+,X_0=x_1,X_T=x_2)$ the probability distribution of the positive   time $t_+$ within the interval $[0,T]$ for a standard Brownian motion $X_t$, starting at $X_0=x_1$ and ending at $X_T=x_2$. For our perturbative expansion it is useful to have this in Laplace variables, namely
\begin{align}
\tilde W^+&(\lambda,s,x_1,x_2)\\
&= \int_{0}^{\infty}\!\!\!\rmd T \int_{0}^{T}\!\!\! \rmd t_+ e^{-s T - \lambda t_+}{\cal P}(t_+,X_0=x_1,X_T=x_2)\nn\ .
\end{align}
We now use the result from the previous section, starting with the special case $x_{1}=x_{2}=0$. The probability distribution for a Brownian that its positive time, up to time $T$, is $t_+$ and that $X_{0}=X_T=0$, i.e.\ the process is a bridge, can be obtained from the discrete case via
\begin{equation}
{\cal P}(t_+,X_T) \rmd t_+ \rmd X_T \Big|_{X_T=0}\underset{\delta t \to 0}{\simeq} \frac{1}{2^n} N^{\rm pos}_{n,k}\ .
\end{equation}
Here $n= \mbox{floor} (T/\delta t)$, $k=\mbox{floor}( t_+/\delta t )$, and   $\delta t$ is the time discretisation step. 
This allows us to relate the generating function \eqref{PostimeBridgeGenerating} to the Laplace transform of the continuous-time distribution $\tilde W^+$ with $x_1=x_2=0$, which we   denote $\tilde W^+(\lambda,s)$, setting  $\nu \to e^{-\delta t \lambda}$, $\rho \to e^{-\delta t s}$ and then taking the limit of $\delta t \to 0$. The measure $\rmd t_+ \rmd B_T$ gives a factor of $\sqrt{2} \delta t^{3/2}$,   c.f.~\Eqref{RandomWalkLimit}. This yields
\begin{align}
\tilde W^+(\lambda,s) \sqrt{2} \delta t^{3/2} &\simeq \tilde p^{\rm pos}(e^{-\delta t \lambda} ,e^{-\delta t s})\delta t^{2 } \nn \\ &\simeq\frac{2 \delta t^2}{\sqrt{1-e^{-2\delta t (s + \lambda)}}+\sqrt{1-e^{-2\delta t s }}}\nn\\
&\simeq \frac{\sqrt{2} \delta t^{3/2}}{\sqrt{\lambda +s}+\sqrt{s}} + \mathcal{O}(\delta t^2)\ .
\end{align}
Thus
\be \label{ContinuousLimit}
\tilde W^+(\lambda,s) = \frac{1}{\sqrt{\lambda +s}+\sqrt{s}}\ .
\ee
From this result for the bridge we   obtain the expression for $\tilde W(\lambda,s,x_1,x_2)$ by distinguishing two cases, see Fig.~\ref{f:W-cases}: 
\begin{figure}
\Fig{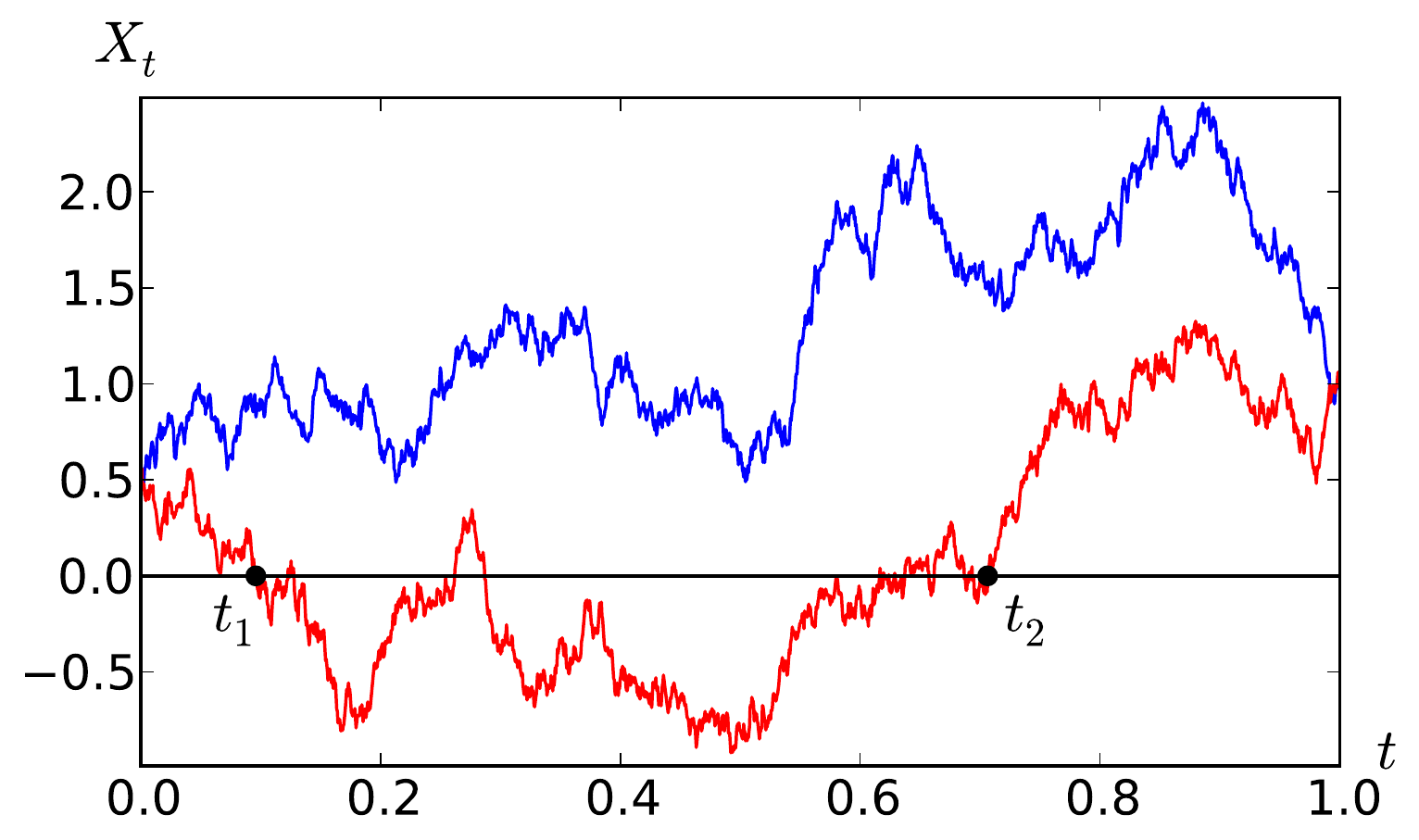}
\caption{In red (bottom curve) a contribution to $\tilde W_1^+(\lambda,s,x_1,x_2)$, where the path reaches 0 at least once (here  for $x_{1}=0.5$ and $x_{2}=1$). In blue  (top curve) the additional contribution  to $\tilde W_2^+(\lambda,s,x_1,x_2)$,  where the path never reaches 0, possible when $x_{1}$ and $x_{2}$ have the same sign (here  for $x_{1}=0.5$ and $x_{2}=1$).}
\label{f:W-cases}
\end{figure}
The first case is when the process changes sign at least once. It can   be decomposed into a constant-sign part (contributing to $t_+$ or not, depending on the signe of $x_1$), a bridge part, and another constant sign part ending in $x_2$. The other case is when the process never changes sign, which corresponds to the survival probability and can be expressed using the method of   images. 

We recall the Laplace transform of this propagator from $x_{1}$ to $x_{2}$, conditioned that the  path has never touched zero \cite{DelormeWiese2016},  
\be \label{Z0+xy}
\tilde P_0^{+}(x_{1} ,x_{2};s) = \frac{e^{-\sqrt{s }|x_1-x_2|} -e^{-\sqrt{s }|x_1+x_2|}}{2\sqrt{s  }} \,\Theta(x_1x_2) \ .
\ee
The {\em normalized} limit  $x_1 \to 0$ is 
\be\label{Z0+x}
\tilde P_0^{+}(x_2;s) = \lim_{x_1\to0} \frac1{x_1}\tilde P_0^{+}(x_{1} ,x_{2};s) = e^{-\sqrt{s }x_2 }\, \Theta(x_2)\ .
\ee
The final result is the sum of two terms,
\begin{equation} \label{PostTimePropagator}
\tilde W^+(\lambda,s,x_1,x_2) = \tilde W_1^+(\lambda,s,x_1,x_2)+ \tilde W_2^+(\lambda,s,x_1,x_2)\ .
\end{equation}
The first contribution   involves a crossing, and is a product of two factors  \eqref{Z0+x} and one factor \eqref{ContinuousLimit},
\begin{align}
&\tilde W_1^+(\lambda,s,x_1,x_2) \nn\\
&\;\;= e^{-\sqrt{s+\lambda\Theta(x_1)}|x_1| } \frac{1}{\sqrt{s+\lambda}+\sqrt{s}}e^{-\sqrt{s+\lambda\Theta(x_2)}|x_2| }\ ,
\end{align}
The $\Theta$ functions in the exponential are understood as follows: If $x_{1}>0$, then $s$ is changed to $s+\lambda$, since this segment contributes both to $T$ and $t_{+}$. In the opposite case $x_{1}<0$, this segment contributes only  to $T$ but not to $t_{+}$, thus $s$ remains unchanged. The same argument applies to the last factor as a function of the sign of $x_{2}$.

The contribution when the walk never changes sign is 
\begin{align}
&\tilde W_2^+(\lambda,s,x_1,x_2)\\
&= \frac{e^{-\sqrt{s+\lambda\Theta(x_1)}|x_1-x_2|} -e^{-\sqrt{s+\lambda\Theta(x_1)}|x_1+x_2|}}{2\sqrt{s + \lambda \Theta(x_1)}} \Theta(x_1x_2)\nn\ .
\end{align}
This is the propagator (\ref{Z0+xy}), with again $s$ shifted to $s+\lambda$ if $x_{1}$, and as a consequence also $x_{2}$, are positive.

The result for $\tilde W^+(\lambda,s,x_1,x_2)$ can also be obtained   by solving the Fokker-Planck equation
\begin{align}
&\!\!\!\partial_{x_2}^2 \tilde W^+(\lambda,s,x_1,x_2) \nn\\
&\quad = \left[s+\lambda \Theta(x_2) \right]\tilde W^+(\lambda,s,x_1,x_2) + \delta(x_1-x_2)\ .
\end{align}
One  verifies that $\tilde W_1^+   + \tilde W_2^+$ is indeed a solution.

As a check, we   consider   Brownian motion starting at $0$ and without any constraint  at the end point, by integrating $\tilde W_+$ over the last variable
\begin{equation}\label{19}
\int_{-\infty}^{\infty} \rmd x\, \tilde W_+(\lambda,s,0,x) =  \frac{1}{\sqrt{s(s+\lambda)}}\ .
\end{equation}
The corresponding probability  distribution for $t_+$   is known as one of the Arcsine laws,
as given in \Eqref{BrownianArcsinLaws}. 
Indeed,  computing the double Laplace transform from this known result yields  \Eqref{19}:
\begin{equation}
\int_{0}^{\infty}\!\!\!\!\!\rmd T \int_{0}^{T}\!\!\!\! \rmd t_+ e^{-s T - \lambda t_+}\frac{1}{\pi \sqrt{t_+(T-t_+)}} =\frac{1}{\sqrt{s(s+\lambda)}}\ .
\end{equation}
For  a Brownian Bridge, i.e.\ $x_1=x_2=0$, we have
\begin{equation}\label{BrownianBridgePosTimeLaplace}
\tilde W^+(\lambda,s,0,0) =\tilde W^+(\lambda,s) 
=\frac{1}{\sqrt{\lambda +s}+\sqrt{s}}\ .
\end{equation}
Let us note  some subtleties. Eq.~(\ref{BrownianBridgePosTimeLaplace})  is the double Laplace transform  of the probability distribution that the Brownian process spends a time $t_+$ in the positive half space {\em and} ends in $0$ at time $T$. If we want to have the condtional probability distribution for $t_+$, knowing that the process is a bridge, we need to divide the result by the probability density to return to $x=0$ at time $T$, which is $(2\sqrt{ \pi T})^{-1}$. The double Laplace transform to compute is then
\begin{equation}
\int_{0}^{\infty}\!\!\rmd T \int_{0}^{T}\!\! \rmd t_+ e^{-s T - \lambda t_+}\frac{1}{T} \frac{1}{2\sqrt{\pi T}} =\frac{1}{\sqrt{\lambda +s}+\sqrt{s}}\ .
\end{equation}
Here $1/T$ is the uniform probability distribution \eqref{BrownianBridgeArcsinLaws} of $t_+$ for a Brownian Bridge, and $(2\sqrt{ \pi T})^{-1}$ is the probability density to return to $0$ at time $T$.
This indeed reproduces Eq.~(\ref{BrownianBridgePosTimeLaplace}).

\subsection{Scale invariance and a useful transformation}
\label{s:SalingTransformation}

The fact that   fBm is a scale invariant (i.e.\ self affine) process implies interesting properties for various distributions. For  $t_+$, and similarly for other temporal observables, the distribution ${\cal P}^T_H(t_+)$ for a fBm process defined on $[0,T]$ (with either a free end-point or a constrained one) takes the   scaling form
\begin{equation}\label{PosTimeDistribScaling}
{\cal P}^T_H(t_+)= \frac{1}{T} \, g\!\left(\vartheta = \frac{t_+}{T}\right)\ .
\end{equation}
Using this, the double Laplace transform of the distribution can be reformulated using a one-variable transformation:
\begin{equation}
\begin{split}
\tilde{\cal P}_H(\lambda,s) &=\int_{0}^{\infty}\rmd T \int_{0}^{T} \rmd t_+ e^{-s T - \lambda t_+} {\cal P}^T_H(t_+)\\
&= \int_{0}^{\infty} \rmd T \int_{0}^{1} \rmd \vartheta\, e^{-T (s + \lambda \vartheta) } g({ \vartheta})\\
&= \frac{1}{s}\int_0^1 \rmd{ \vartheta} \frac{g(\vartheta)}{1+ \frac{\lambda}{s} \vartheta}  = \frac{1}{s} \bar g\!\left(\kappa = \frac{\lambda}{s}\right).
\end{split}
\end{equation}\\
The scaling function $g(\vartheta)$ encoding the distribution ${\cal P}^T_H(t_+)$, and the scaling function $\bar g(\kappa)$ encoding its double Laplace transform $\tilde{P}(\lambda,s)$, are  related by a simple integral transform  which we denote $\mathcal{K}_1$,
\begin{equation}
 \mathcal{K}_1\!\left[g\right]\!(\kappa):=  \int_0^1 \rmd \vartheta\, \frac{ g(\vartheta)}{1+ \kappa \vartheta} = \bar g(\kappa) \ .
\end{equation}
For the case of interest, a fBm bridge of lenght $T$, this relation is more complicated since we can not compute directly the double Laplace transform of ${\cal P}^{\rm bridge}_H(t_+)$, but only the transform of an unnormalised distribution, which we   write $Z^N(T) {\cal P}^{\rm bridge}_H(t_+)$. As we will see, the normalisation factor  $Z^N(T)$, which is the  probability density to return to $0$ at time $T$, is a power law, 
\begin{equation}
Z^N(T)= C \,T^{\alpha-1}\ ,
\end{equation}
with some constant $C$. In this case, the double Laplace transform of the unnormalised distribution is computed as
\begin{align}\label{KtransformGeneral} \nn
\int_{0}^{\infty}\rmd T \int_{0}^{T}& \rmd t_+ e^{-s T - \lambda t_+} CT^{\alpha -1 }{\cal P}^{\rm bridge}_H(t_+)\\  \nn
&= \int_{0}^{1} \rmd { \vartheta} \int_{0}^{\infty} \rmd T \,CT^{\alpha-1} e^{-T (s + \lambda \vartheta) } g(\vartheta)\\ \nn
&= \frac{C\,\Gamma({ \alpha} )}{s^{ \alpha }}\int_0^1 \rmd \vartheta \frac{g(\vartheta)}{\left(1+ \frac{\lambda}{s} \vartheta\right)^{\alpha}} \\
&\overset{!}{=} \frac{C\,\Gamma({\alpha })}{s^{ \alpha}}\mathcal{K}_{ \alpha }\!\left[g\right]\!\left(\kappa =\frac{\lambda}{s}\right)\ .
\end{align}
Here we   generalised the $\mathcal{K}$ transform to another exponent, 
\begin{equation}\label{Ktransform}
\mathcal{K}_{\alpha}[g](\kappa):=\int_0^1 \rmd \vartheta\, \frac{ g(\vartheta)}{(1+\kappa \vartheta)^{\alpha}}\ .
\end{equation}
\\
If $\bar g (\kappa) = \mathcal{K}_{\alpha}[g](\kappa)$ is the $\mathcal{K}_{\alpha}$ transform of a function $g(\vartheta)$ normalised to unity, then   $\bar g(\kappa) \to 1$ for $\kappa \to 0$. If further $g(\vartheta)$ is  time-reversal symmetric,   $g(\vartheta) = g(1-\vartheta)$ for $\vartheta \in [0,1]$, then the function $\bar g(\kappa)$ has the   symmetry
\begin{equation}
\bar g(\kappa)= \frac{1}{(1+\kappa)^{\alpha}} \bar{g}\!\left(- \frac{\kappa}{1+\kappa}\right)\ .
\end{equation}
\subsection{FBm bridge with  $H=\frac12+\epsilon $}
The path-integral approach presented in Section \ref{s:Perturbative} yields an  expression for the (unnormalised) density distribution of $t_+$  for a bridge,
\begin{equation}
\begin{split}
Z^{\rm pos}&(t_+,T) =\\
& \int_{X_0=0}^{X_T=0} \mathcal{D}[X] \,\delta\!\left(\int_0^T\!\!\! \rmd t\, \Theta(X_t) - t_+\right) e^{-S[X]}\ .
\end{split}
\end{equation}
It is useful to consider its double Laplace transform ($ T \to s$ and $t_+ \to \lambda$), which we denote with a tilde
\begin{equation}\label{PathIntegralPosTimeLaplace}
\begin{split}
\tilde{Z}^{\rm pos}&(\lambda,s) =\\
&\!\!\!\int_0^{\infty}\!\!\! \rmd T\, e^{- s T} \int_{X_0=0}^{X_T=0} \mathcal{D}[X]\, e^{-S[X] -\lambda \int_0^T \rmd t\, \Theta(X_t) } \ .
\end{split}
\end{equation}
Using the $\varepsilon$-expansion  \eqref{ActionExpansion} for the action, we   compute this perturbatively, expanding around   Brownian motion. The resulting    series in $\varepsilon$ has the form
\begin{equation}
\tilde{Z}^{\rm pos}(\lambda,s)=\tilde{Z}^{\rm pos}_0(\lambda,s) + \varepsilon \tilde{Z}^{\rm pos}_1(\lambda,s) + \mathcal{O}(\varepsilon^2)\ .
\end{equation}
The first term of this expansion, the   result for   Brownian motion, is as in \Eqref{BrownianBridgePosTimeLaplace}  obtained   from the propagator $\tilde W^+$, 
\begin{equation}\label{ZposOrdre0}
\tilde Z^{\rm pos}_0(\lambda,s)=\tilde W^+(\lambda,s)=\frac{1}{\sqrt{s}} \frac{1}{\sqrt{1+\kappa}+1}=\frac{\bar{g}_0(\kappa)}{{ 2}\sqrt{s}}\ .
\end{equation}
Here we   denoted
\begin{equation}
\bar g_0(\kappa)=\int_0^1 \!\rmd { \vartheta} \frac{ g_0({ \vartheta})}{\sqrt{1+\kappa { \vartheta}}}= \frac{2}{\sqrt{1+\kappa}+1}\ .
\end{equation}
This can be inverted to  
\be
g_0(\vartheta) = 1\ .
\ee
This reproduces the known result that the probability distribution \eqref{BrownianBridgeArcsinLaws} for a Brownian bridge is uniform \cite{Levy1940,NikitinOrsingher2000}.

To compute the order-$\varepsilon$ term $\tilde{Z}_1^{\rm pos}(\lambda,s)$, we   use the same diagrammatic rules as  in Ref.~\cite{DelormeWiese2016}, Section III~D. These rules are easily expressed in Laplace variables, which is why we compute the expansion of $\tilde{Z}^{\rm pos}(\lambda,s)$. The first order-$\varepsilon$ correction comes from the non-local interaction in the action, given in the second line of \Eqref{ActionExpansion}, and can be written as
\begin{widetext}
\begin{equation}\label{Zpos1Adiag}
\tilde{Z}_{1 \rm A}^{\rm pos}(\lambda,s)= {2} \int_0^{\Lambda}\!\! \rmd y\int_{-\infty}^{\infty}\!\!\rmd x_1 \int_{-\infty}^{\infty}\!\!\rmd x_2\;\tilde W^+(\lambda,s,0,x_1)\,\partial_{x_1}\! \tilde W^+(\lambda,s+y,x_1,x_2)\,\partial_{x_2}\!\tilde W^+(\lambda,s,x_2,0) \ .
\end{equation}
As explained in Ref.~\cite{DelormeWiese2016}, the large-$y$ cutoff $\Lambda$, which is necessary as the integral is logarithmically divergent, is linked to the short-time (UV) regularisation $\tau$ introduced in \Eqref{ActionExpansion} by
$\Lambda = e^{-\gamma_{\rm E}}/ \tau$. Performing the integrations over space variables and over $y$, and after some simplifications, we obtain
\begin{equation}\label{Zpos1A}
\tilde{Z}^{\rm pos}_{1 \rm A}(\lambda,s)= \frac{1}{ \sqrt{s}}\left[ \left(\frac{4}{\sqrt{\kappa+1}}+4\right)\!\log \left(\sqrt{\kappa+1}+1\right)-\frac{2 \kappa +2+\sqrt{\kappa +1}}{ \kappa }\log (\kappa +1)+\frac{\log (s \tau)+7-7 \log (4)+\gamma_{\rm E}}{ \sqrt{\kappa +1}+1}\right]\ .
\end{equation}
\end{widetext}
We have expressed the result in terms of the dimensionless variable $\kappa = \lambda /s$. 
The second order-$\varepsilon$ correction comes from the rescaling of the diffusive constant, c.f.~Eq.~\eqref{DiffusConstant}. It is computed by rescaling   $T$ in the  result for the Brownian, setting $T \to D_{\varepsilon,\tau} T$. In Laplace variables, this is equivalent to writing
\begin{equation}
\tilde Z_0^{\rm pos}(\lambda,s) \to \frac{1}{D_{\varepsilon,\tau}} \tilde Z_0^{\rm pos}\!\left(\frac{\lambda}{D_{\varepsilon,\tau}},\frac{s}{D_{\varepsilon,\tau}}\right)\ .
\end{equation}
Extracting the order-$\varepsilon$ term gives
\begin{equation}
\tilde Z^{\rm pos}_{1B}(\lambda,s) = - \frac{1+\ln(\tau)}{2\sqrt{s}} \frac{2}{\sqrt{1+\kappa}+1}\ .
\end{equation}
Resumming all order-$\varepsilon$ corrections,
\begin{equation}
\tilde Z_1^{\rm pos}(\lambda, s) = \tilde Z_{1A}^{\rm pos}(\lambda, s)+\tilde Z_{1B}^{\rm pos}(\lambda, s)\ ,
\end{equation} the $\tau$ dependence vanishes. The $\ln(s)$ term in \Eqref{Zpos1A} is proportional to $\bar g_0(\kappa)$, such that we can recast it as an order-$\varepsilon$ correction to the exponent of the prefactor: $s^{-1/2} \to s^{H-1} + \mathcal{O}(\varepsilon^2)$. This allows us to write the path integral \eqref{PathIntegralPosTimeLaplace} in the form
\begin{equation}\label{39}
\tilde Z^{\rm pos}(\lambda,s) = \frac{\Gamma(1-H)}{2 \sqrt{\pi} s^{1-H}}\Big[\bar{g}_0(\kappa)+\varepsilon \bar g_1^{\rm pos}(\kappa)\Big]+\mathcal{O}(\varepsilon^2)\ .
\end{equation}
With this choice of  prefactor, the constant $C$ in   \Eqref{KtransformGeneral} is   $C=(2 \sqrt{\pi})^{-1}$, and $\bar{g}_1^{\rm pos}(\kappa)$ is
\begin{align}
\bar g_1^{\rm pos}(\kappa)=&\;8\left(\frac{1}{\sqrt{\kappa+1}}+1\right)\!\log \left(\sqrt{\kappa+1}+1\right)\\
&\!-2\frac{2 \kappa +2+\sqrt{\kappa +1}}{ \kappa }\log (\kappa +1)+4\frac{3-4 \log (4)}{ \sqrt{\kappa +1}+1}\nn\ .
\end{align}
We recall that this function contains contributions from $\tilde{Z}^{\rm pos}_{1 \rm A}$, $\tilde{Z}^{\rm pos}_{1 \rm B}$ and the expansion of 
$\frac{1}{\sqrt{\pi}}\Gamma\!\left(\frac12 - \varepsilon\right) = 1+\varepsilon\big[\gamma_{\rm E}+\ln(4)\big]+\mathcal{O}(\varepsilon^2)$, due to the choice of normalisation in \Eqref{39}.

We know that the distribution of the positive time has the form given in Eq.~\eqref{PosTimeDistribScaling}. After expanding it in $\varepsilon$ it gives
\begin{equation}\label{PosTimeDistribExpansion}
{\cal P}^{\rm bridge}_{H=\frac12+\varepsilon}(t_+)=\frac{1}{T}\Big[g_0(\vartheta)+ \varepsilon g_1^{\rm pos}(\vartheta)\Big] + \mathcal{O}(\varepsilon^2)\ ,
\end{equation}
where, as before, ${ \vartheta} = t_+/T$.

We have seen in Section \ref{s:SalingTransformation} that the scaling functions $g(\vartheta)$ and $\bar g(\kappa)$ are related via the $\mathcal{K}_{1-H}$ transform, where the index of the transformation is fixed by the prefactor $s^{H-1}$ in \Eqref{39}.

Expanding \textit{w.r.t.} $\varepsilon$ in the definition of the $\mathcal{K}$ transform gives 
\begin{align} \nn
\bar{g}(\kappa)&= \int_0^1\!\!\rmd \vartheta \frac{1}{(1+\kappa \vartheta)^{\frac12 -\varepsilon}}g(\vartheta)\\
&=\int_0^1\!\!\rmd \vartheta \frac{1+\varepsilon \log(1+\kappa \vartheta)}{\sqrt{1+\kappa \vartheta}}\left[g_0(\vartheta)+\varepsilon g_1(\vartheta)\right]+\mathcal{O}(\varepsilon^2)\nn \\
&=\bar g_0(\kappa)+\varepsilon \int_0^1 \rmd \vartheta  \frac{\big[g_1(\vartheta)+g_0(\vartheta)\ln(1+\kappa \vartheta)\big]}{\sqrt {1+\kappa \vartheta}} \nn\\
&~~+\mathcal{O}(\varepsilon^2)
\end{align}
The order-$\varepsilon$ correction $g_1(\vartheta)$ that we are looking for is then given by 
\begin{equation}\label{g1Inverse}
g_1(\vartheta)= \mathcal{K}_{\frac12}^{-1}\left[\bar g_1(\kappa)-\bar g_{0,1}(\kappa)\right]\ ,
\end{equation}
where we have defined
\begin{equation}
\begin{split}
\bar g_{0,1}(\kappa) &= \int_0^1 \!\! \rmd \vartheta\, \frac{\ln(1+\kappa \vartheta)}{\sqrt{1+\kappa \vartheta}} g_0(\vartheta)\\&= \frac{2}{\kappa} \left\{2+\sqrt{1+\kappa}\big[\ln(\kappa+1)-2\big]\right\}\ .
\end{split}
\end{equation}
This contribution is valid both for  $t_+$ and $t_{\rm max}$, since both observables have the same distribution at order zero, and both   have the same power law from scaling. 

We now have to deal with the inverse   $\mathcal{K}_{\half}$ transform  in \Eqref{g1Inverse}. This is linked to the Abel transform, on which   details  are given in Appendix \ref{a:Abel}. The final result for the order-$\varepsilon$ correction is
\begin{align}\label{ResultPosTimeOrdre1}
g_1^{\rm pos}(\vartheta)=4\bigg[&2-\frac{1}{\sqrt{\vartheta}+1}+\ln\!\left(\frac{\sqrt{\vartheta }+1}{4 \sqrt{\vartheta }}\right)\\
&-\frac{1}{\sqrt{1-\vartheta}+1}+\ln\!\left(\frac{\sqrt{1-\vartheta }+1}{4 \sqrt{1-\vartheta }}\right)\bigg]\nn\ .
\end{align}
We can check that the integral of $g_1^{\rm pos}(\vartheta)$ over $[0,1]$ vanishes, such that \Eqref{PosTimeDistribExpansion} is correctly normalised at order $\varepsilon$. We   also checked that by computing  numerically the  $\mathcal{K}_{1/2}$ transform of this result  reproduces $\bar g_1^{\rm pos}(\kappa)-\bar g_{0,1}(\kappa)$ with excellent precision.

Close to the boundary,  the   asymptotics is
\begin{equation}
g^{\rm pos}_1(\vartheta) \underset{\vartheta \to 0,1}{\simeq} -2 \ln(\vartheta)-2\ln(1-\vartheta) \ .
\end{equation}
This asymptotics can be recast into a power law   consistent with scaling. The distribution of $t_+$ for a fBm bridge with $H=\frac12 + \varepsilon$ can then be written as 
\begin{equation}\label{fBmPosTimeDistrib}
{\cal P}^{\rm bridge}_{H=\frac12 + \varepsilon}(t_+)= \frac{\exp\!\Big(\varepsilon \left[ \mathcal{F}^{\rm pos}(\vartheta)-4 \right]\Big)}{T[\vartheta(1-\vartheta)]^{2H-1}}+\mathcal{O}(\varepsilon^2)\ .
\end{equation}
The scaling function $\mathcal{F}^{\rm pos}(\vartheta)$ has by definition vanishing integral,   and is given by
\begin{align}\label{PostimeScalingF}
\mathcal{F}^{\rm pos}(\vartheta)
= 4 \Bigg[ 3 & -\frac{1}{\sqrt{\vartheta}+1}+\ln\!\left(\frac{\sqrt{\vartheta }+1}{4}\right) \nn\\
&  -\frac{1}{\sqrt{1-\vartheta}+1}+\ln\!\left(\frac{\sqrt{\vartheta }+1}{4}\right)  \Bigg]\ .
\end{align}

\subsection{Numerical results}
\begin{figure}[t]
\Fig{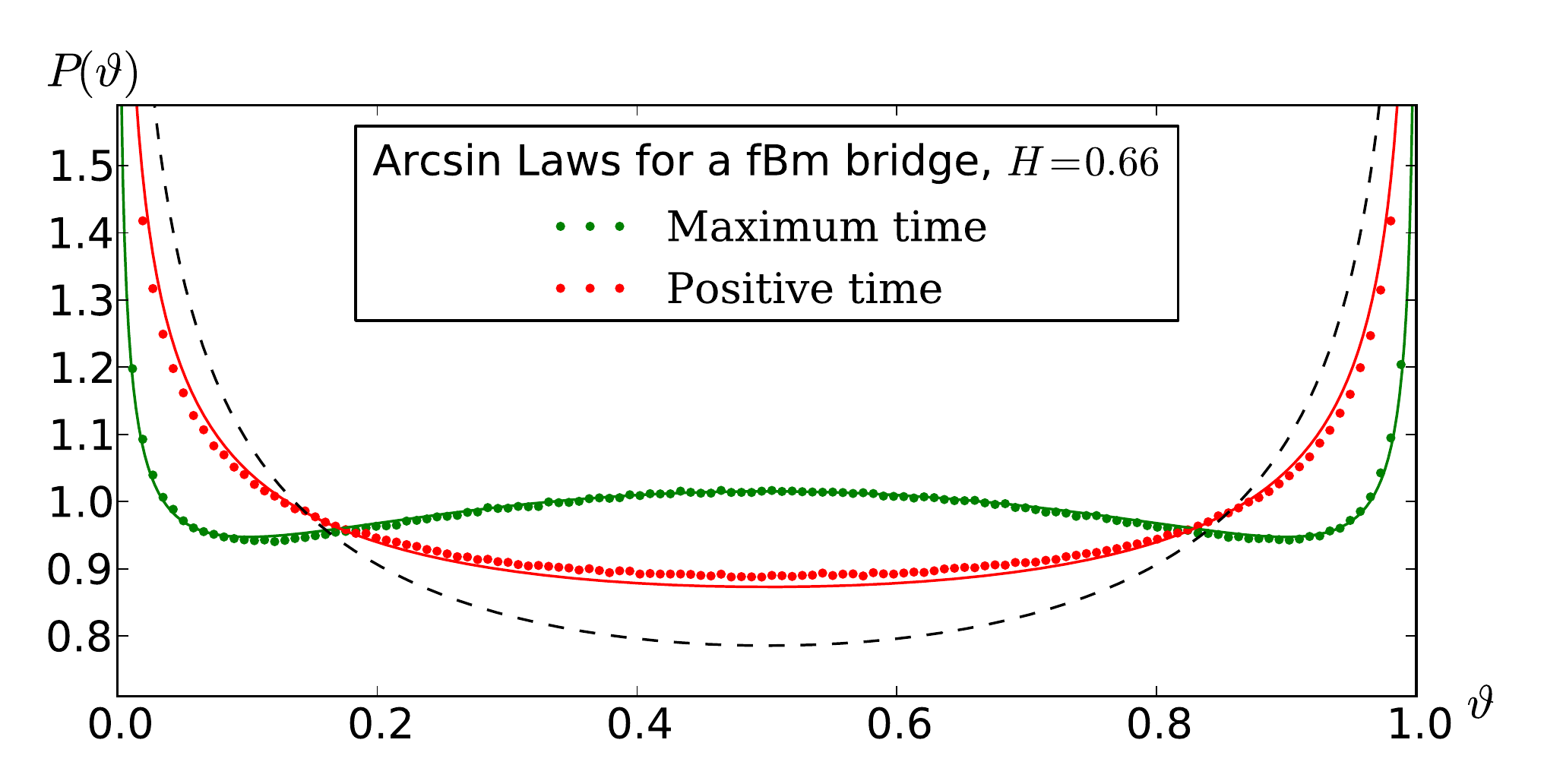}
\caption{Comparison of the two ``Arcsine laws'' for a fBm bridge with Hurst exponent $H=0.66$. Dots represent the distribution extracted from numerical simulations, the plain lines represent the analytical result at order $\varepsilon$ given in Eqs.~\eqref{fBmPosTimeDistrib} and \eqref{fBmTmaxDistrib}, and the dashed line is the scaling form (identical for both observables).}\label{ArcsinLawsFigure}
\end{figure}

\begin{figure*}[t]
\Fig{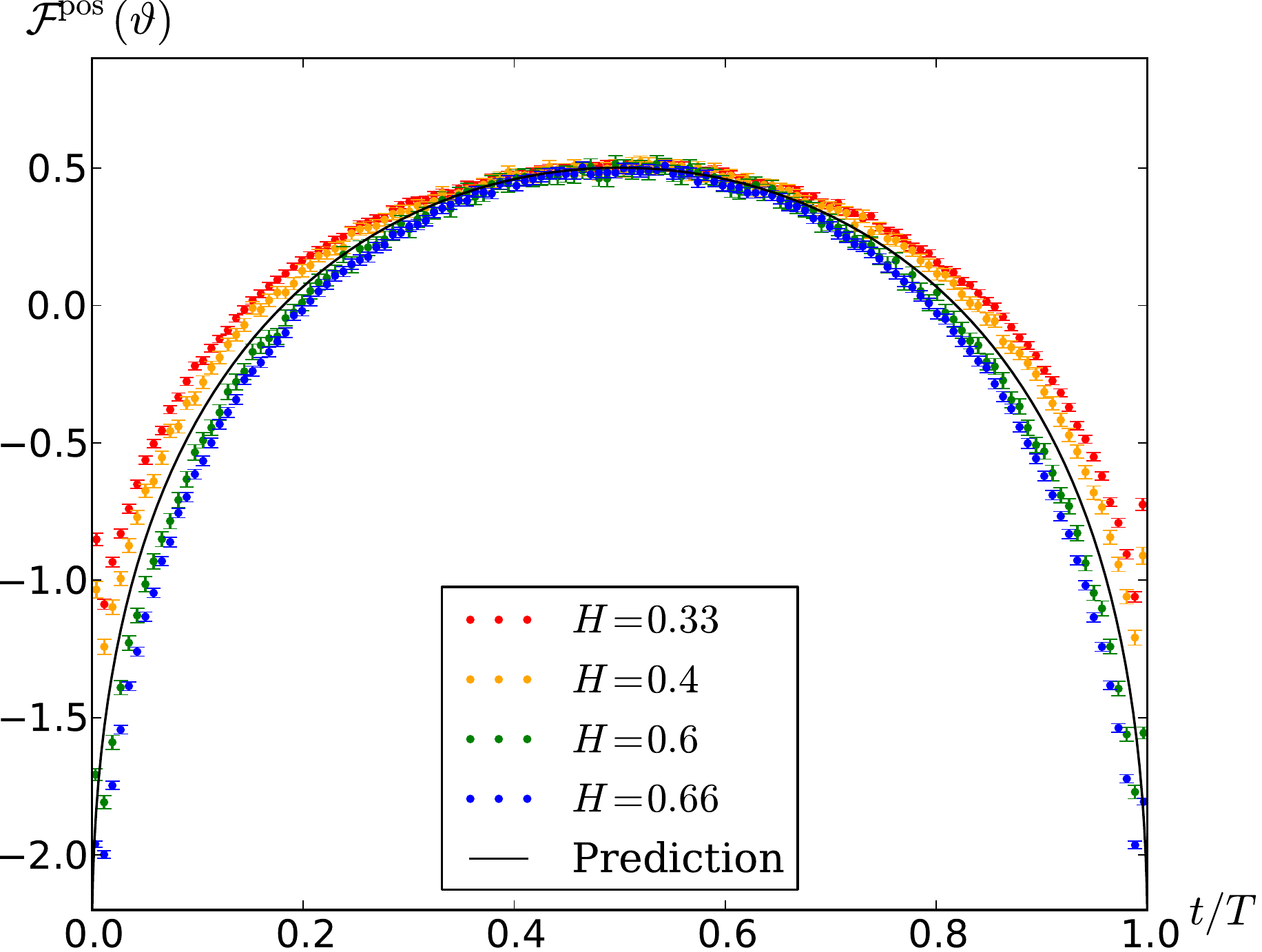}~~~~\Fig{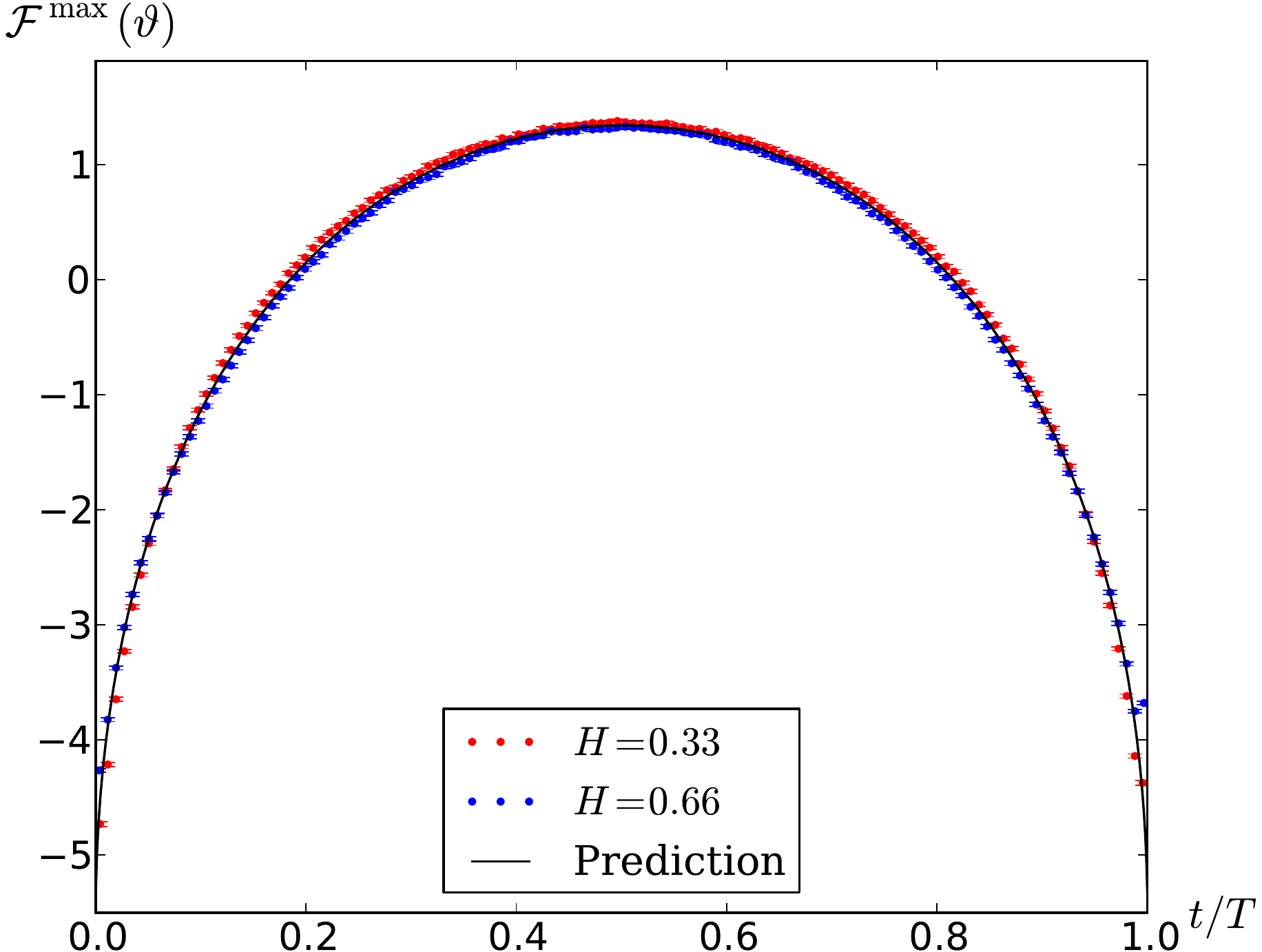}
\caption
{Left: Numerical estimation of the scaling function $\mathcal{F}^{\rm pos}(\vartheta)$, from top to bottom for $H=0.33$ (red dots), $H=0.4$ (orange dots), $H=0.6$ (green dots), and $H=0.66$ (blue dots), compared to the analytical result  given in \Eqref{PostimeScalingF} (plane line). Right: {\it ibid} for  $\mathcal{F}^{\rm max}(\vartheta)$ for $H=0.33$ (blue dots, bottom) and $H=0.66$ (red dots, top), the analytical result (plane line) is given in \Eqref{TmaxScalingF}. For both plots, and for each value of $H$, the statistics is done with $5 \times 10^{6}$ sampled paths, discretized with $N=2^{12}$ points.}
\label{PostimeFBmBridgeFig}
\end{figure*}

To test our analytical predictions, we compare them to results from numerical simulations. As in Ref.\ \cite{DelormeWiese2016}, we construct a large number of fBm paths using the Davis and Harte procedure, c.f.~Ref.~\cite{DiekerPhD}  for details on the numerical method. From these samples, we   construct a  numerical estimation ${\cal P}^{\rm bridge}_{H}(t_+)$ of the distribution of $t_+$ for various values of $H$,   choosing $T=1$. 
This is shown on Fig.~\ref{ArcsinLawsFigure}, where results for the distributions of both  $t_{+}$ and $t_{{\rm max}}$ are given. 
To compare to the analytical result \eqref{PostimeScalingF}, we extract $\mathcal{F}^{\rm pos}_{\rm num}$ from these distributions, using
\begin{equation}
\mathcal{F}^{\rm pos}_{\rm num}(\vartheta) = \frac{1}{\varepsilon} \ln\!\Big(T [\vartheta(1-\vartheta)]^{2H-1}{\cal P}^{\rm bridge}_{H=\frac12 + \varepsilon}(\vartheta)\Big)\ .
\end{equation}
As is shown in Fig.\ \ref{PostimeFBmBridgeFig}  (left),   when $\varepsilon \to 0$, $\mathcal{F}^{\rm pos}_{\rm num}(\vartheta)$ converges to $\mathcal{F}^{\rm pos}(\vartheta)$. The deviation being antisymmetric in $\varepsilon$   strongly suggests  that there is an order-$\varepsilon^2$ correction to the distribution of $t_+$, which we did not calculate here.

\section{Extremum of fBm Bridges}\label{s:ExtremeValues}
In Ref.~\cite{DelormeWiese2016}, a general formula was derived for the path integral over fBm paths $X_t$ starting at $m_1$, going to $x_0 \approx 0$ at time $t_1$ and ending in $m_2$ at time $t_1+t_2=T$, while staying positive, $X_t>0$ for all $t \in [0,T]$. This quantity, denoted $Z^+(m_1,t_1;x_0;m_2,t_2)$, is the first-order term in an $\varepsilon$ expansion.  It was used \cite{DelormeWiese2016} to derived results about extremal properties of   fBm in the unconstrained case: Both the distribution of the  maximum $m$, and the time  $t_{\rm max}$ when this maximum is achieved,  were computed, as well as their joint distribution. Comparison  to numerical simulations showed that this result is of great precision for small $\varepsilon$, and of good precision for larger vales of $\varepsilon$. 

Here we apply these results to fBm bridges. The general result for $Z^+(m_1,t_1;x_0;m_2,t_2)$, restricted to $m_1=m_2=m$, and choosing $t_1+t_2 =T$, immediately gives the joint distribution of the maximum $m$, and the time $t_{\rm max}=t_1$ when  this maximum is attained. 
In a second step, we can then integrate over $t_1$ at $T$ fixed, or over $m$ at $t_1$ and $t_2$ fixed, to obtain the distributions of $m$ and $t_{\rm max}$. 

We will finally  rederive these results in a simpler way, taking advantage of the scaling transformations introduced in section \ref{s:SalingTransformation}.

\subsection{Distribution of the time to reach the maximum}

Starting with Eq.~(44) of Ref.~\cite{DelormeWiese2016} and following the   procedure   in its section IV.C, we   express the probability     for $t_{\rm max}$, denoted ${\cal P}^{\rm bridge}_{H}(t_{\rm max})$, as
\begin{equation}\label{PtmaxZ}
{\cal P}^{\rm bridge}_{H}(t_{\rm max})= \frac{1}{Z^N(T)}\int_{0}^{\infty}\!\!\!\rmd m\, Z^+(m,t;x_0;m,T-t)\ .
\end{equation}\\
 The integral over $m$  accounts for all possible values of the maximum. 
 $Z^N(T)$ is a normalisation factor such that the integral over $t_{\rm max}$ of ${\cal P}^{\rm bridge}_{H}(t_{\rm max})$ is normalised to unity,
\begin{equation}
\begin{split}
Z^N(T)&=\int_{0}^{T}\!\!\rmd t\int_{0}^{\infty}\!\!\!\rmd m\, Z^+(m,t;x_0;m,T-t)\\
&= \frac{x_{0}^{ 2-4\varepsilon}}{\sqrt{4 \pi}}(1+ \varepsilon C_1) + \mathcal{O}(\varepsilon^2)\ .
\end{split}
\end{equation}
The constant $C_1$ can be computed from $Z^+$, but it is equivalent to   require that the order-$\varepsilon$  term in \Eqref{PtmaxZ} does not change the normalisation, such that the distribution ${\cal P}^{\rm bridge}_{H}(t_{\rm max})$ remains normalised to one.

Expanding the distribution of $t_{\rm max}$ in the same way as for \Eqref{PosTimeDistribExpansion},   the order-$\varepsilon$ term becomes, setting again $\vartheta =  t_{\rm max} /T$, and $T=1$
\begin{align}\label{CorrectionTmaxDistrib}
& g_1^{\rm max}(\vartheta)= 2 \sqrt{\pi}\int_{0}^{\infty}\!\!\!\rmd m\, \big[ Z_1^+(m,\vartheta ;x_0;m,1-\vartheta) \nn\\
& ~~~~~~~~~~~~~~~~~- C_1  Z_0^+(m,\vartheta ;x_0;m,1-\vartheta)\big]  \nn\\
&\quad =\,2 \Big[6 (\sqrt{1-\vartheta}+ \sqrt{\vartheta})-3 \vartheta \log (1-\vartheta)-3(1-\vartheta) \ln (\vartheta)\nn\\
&\qquad~~~~~+(4-3 \vartheta) \log (2-\vartheta)+(3 \vartheta+1) \log (\vartheta+1)\nn\\
&\qquad~~~~~+(6 \vartheta-4) \text{arcth}(\sqrt{1-\vartheta})+(2-6 \vartheta) \text{arcth}(\sqrt{\vartheta})\nn\\
&\qquad~~~~~  -8 -4 \ln (2)\Big]\ .
\end{align}
This result will be checked    from Eq.~(\ref{ZplusJoint}) given below. 
Demanding that  $g_1^{\rm max}(\vartheta)$ has integral zero fixed the constant $C_1$  to $C_1=4\ln(2)- \gamma_{\rm E}$.

Close to the  boundary, the correction has the same asymptotics as in the calculation for $t_+$, namely
\begin{equation}
g_1^{\rm max}(\vartheta) \underset{\vartheta \to 0,1}{\simeq} -2 \ln(\vartheta)-2\ln(1-\vartheta)\ ,
\end{equation}
which indicates the same change in the power-law behaviour of ${\cal P}^{\rm bridge}_{H}(t_{\rm max})$. 
Again taking an exponential resummation of the order-$\varepsilon$ correction, we obtain a formula similar to \Eqref{fBmPosTimeDistrib},    but with a different scaling function $\mathcal{F}^{\rm max}(\vartheta)$, 
\begin{equation}\label{fBmTmaxDistrib}
{\cal P}^{\rm bridge}_{H=\frac12 + \varepsilon}(t_{\rm max}) = \frac{\exp\!\Big(\varepsilon \big[ \mathcal{F}^{\rm max}(\vartheta)-4\big]\Big)}{T[\vartheta(1-\vartheta)]^{2H-1}}+\mathcal{O}(\varepsilon^2)\ .
\end{equation}
 $\mathcal{F}^{\rm max}(\vartheta)$ is a bounded function of $\vartheta \in [0,1]$ and can be expressed from \Eqref{CorrectionTmaxDistrib} as
\begin{equation}\label{TmaxScalingF}
\mathcal{F}^{\rm max}(\vartheta)= g_1^{\rm max}(\vartheta)+2 \ln\!\big( \vartheta (1-\vartheta)\big) +4\ .
\end{equation}
The constant $4$ was added in \Eqref{TmaxScalingF} and subtracted in \Eqref{fBmTmaxDistrib} to have $\int_0^1\rmd \vartheta \, g_{1}^{\rm max}(\vartheta) =\int_0^1\rmd \vartheta \, {\cal F}^{\rm max}(\vartheta) =0$.

The two distributions, for $t_+$ and $t_{\rm max}$, at order $\varepsilon$ are plotted in Fig.~\ref{ArcsinLawsFigure}.  While both functions have the same power-law behavior for $\vartheta$ close to 0 or 1, their difference is   clearly visible.
The result (\ref{TmaxScalingF}) for $\mathcal{F}^{\rm max}(\vartheta)$ is compared with great precision to numerical simulations on figure \ref{PostimeFBmBridgeFig} (right).

\subsection{The maximum-value distribution}
Similarly to the distribution of $t_{\rm max}$, the distribution of the maximum value $m= \max_{t \in [0,T]} X_t$ can be expressed from the result for $Z^+$ given in Eq.~(44) of Ref.~\cite{DelormeWiese2016}:
\begin{equation}
{\cal P}^{\rm bridge}_{H}(m)= \frac{1}{Z^N(T)}\int_{0}^{T}\!\!\!\rmd t\, Z^+(m,t;x_0;m,T-t)\ .
\end{equation}
This calculation is rather cumbersome, but it is   possible to give a simpler derivation, where we do not constrain paths to go close to the boundary, but construct ${\cal P}^{\rm bridge}_{H}(m)$ by taking a derivative of its cumulative distribution, the survival probability, conditioned such that the end point of the process is the same as the starting point. In this framework, the order-$\varepsilon$ correction to ${\cal P}^{\rm bridge}_{H}(m)$ can,  due to the non-local term in the action \eqref{ActionExpansion},  be expressed in Laplace variables ($T\to s$) using the diagrammatic rules of Ref.~\cite{DelormeWiese2016}.  The integrals to be computed are 
\begin{widetext}\checked
\begin{align}\label{Z1A_Laplace}
\tilde Z_{1A}^{\rm max}(m,s)&=2\partial_m \int_0^{\Lambda} \!\! \rmd y \int_{x_1,x_2>0}\!\!\!\tilde P_0^+(m,x_1;s)\,\partial_{x_1}\tilde P_0^+(x_1,x_2;s+y)\,\partial_{x_2}\tilde P_0^+(x_2,m;s)\\
& =2(a+1) e^{2 a} \,\text{Ei}(-4 a)-2\,\text{Ei}(-2 a) 
+2e^{-2 a} \left\{a \left[\log\!\left(\frac{m^2}{4 \tau }\right)-\ln (a)-1\right] +\log\!\left(\frac{2 \tau }{m^2}\right) - \gamma_{\rm E} \right\}\nn\ ,
\end{align}
where $a:=\sqrt{s} m$ is a dimensionless variable, $\Lambda = e^{-\gamma_{\rm E}}/ \tau$, and the propagator $\tilde P_0^+(x_1,x_2;s)$ is defined in \Eqref{Z0+xy}. To deal with the inverse Laplace transform, we   use   formulas (G10) and (G11) derived in Ref.~\cite{DelormeWiese2016}, plus similar formulas collected in appendix \ref{a:inv-Lap-trafos}.
The final result for the correction after the inverse Laplace transformation is \checked
\begin{equation}
Z_{1A}^{\rm max}(m,T)=\frac{z e^{-z^2}}{\sqrt{\pi}T}  \bigg\{ 2 z \sqrt{\pi } e^{z^2} \text{erfc}(z)  + 4  (1-z^2 ) \mathcal{J} \!\left(z^2\right)+ 2z^2 \left[ \log\!\left(\frac{T z^2}{\tau }\right)+ \gamma_{\rm E} -1\right]+\log\!\left(\frac{\tau ^3}{T^3 z^8}\right)-4 \gamma_{\rm E} +1\bigg\}\ .
\end{equation}
\end{widetext}
We   introduced the scaling variable $
z:=  m/{\sqrt{T}}$. The special function $\mathcal{J}$   defined in Ref.~\cite{DelormeWiese2016} is
\begin{equation}
\mathcal{J}(x)=\frac{1}{2} \pi  \text{erfi}\left(\sqrt{x}\right)-x \, _2F_2\!\left(1,1;\frac{3}{2},2;x\right)\ .
\end{equation}
For a  Brownian bridge we have
\begin{figure*}[t]
\centerline{\fig{5.9cm}{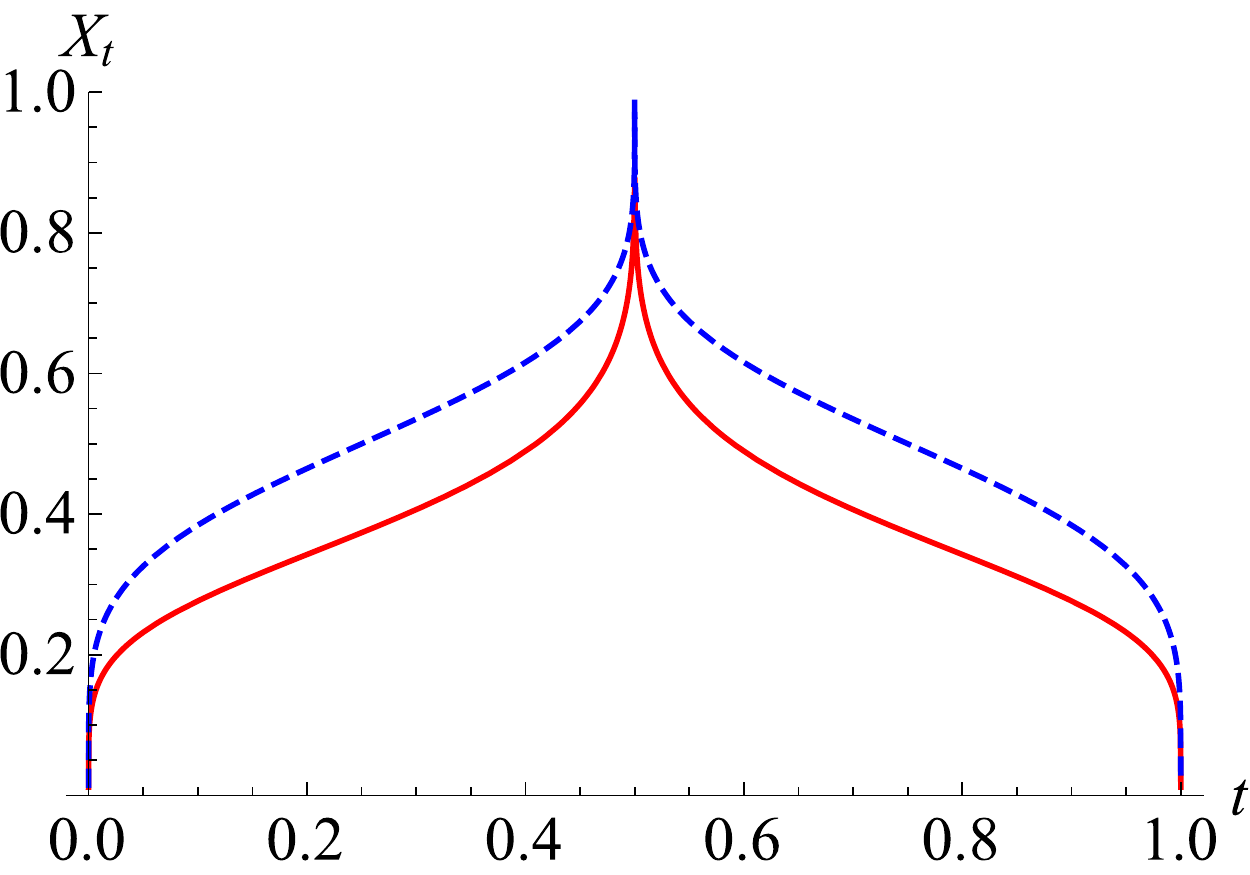}
\fig{5.9cm}{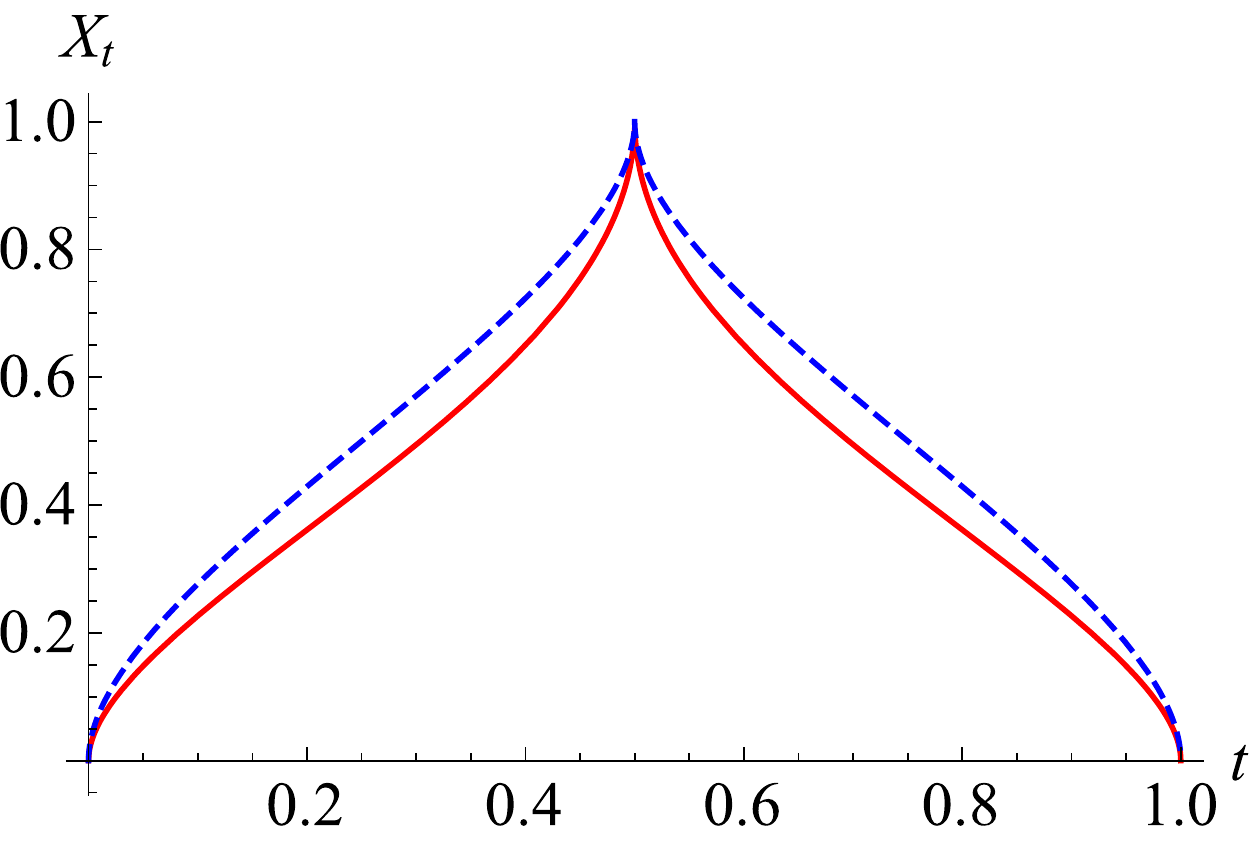}
\fig{5.9cm}{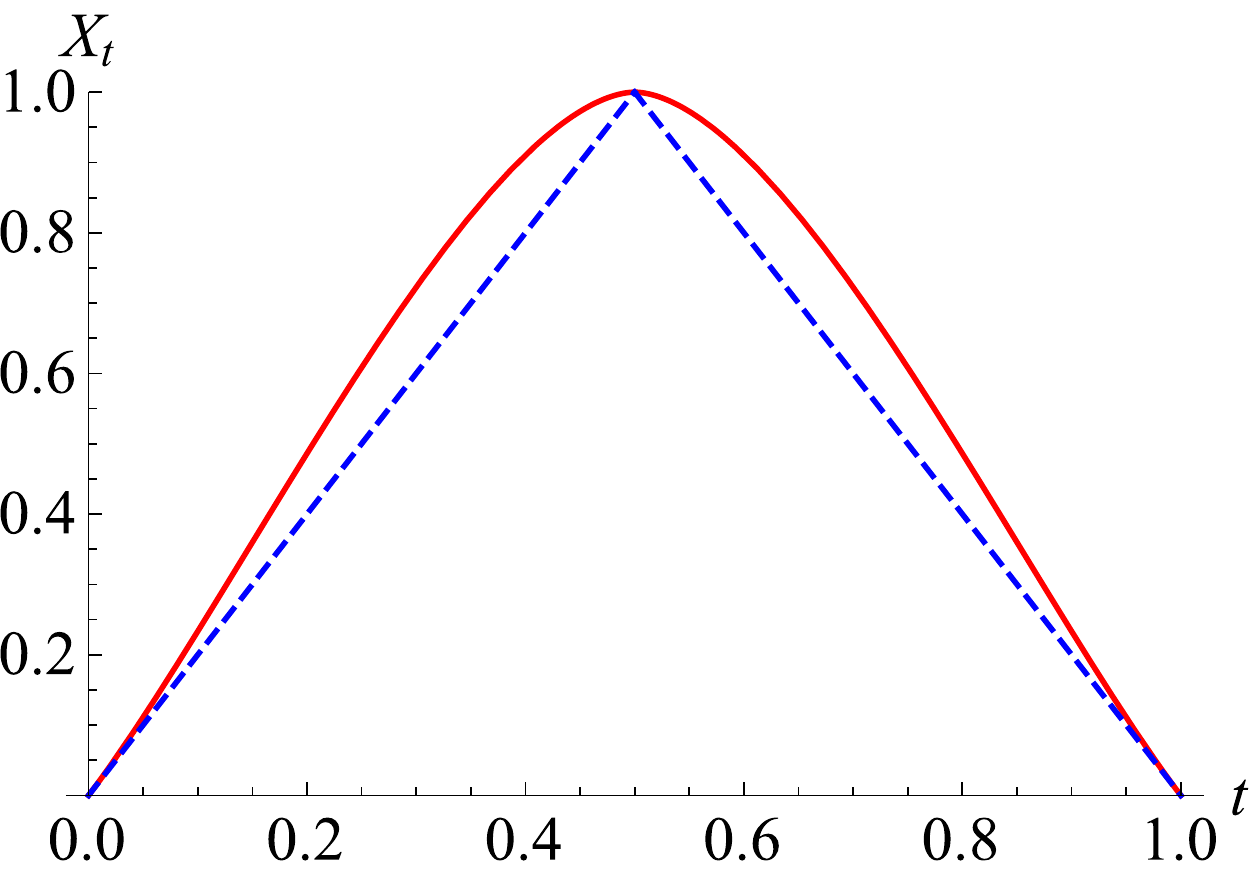}}
\caption{Plain red line: optimal paths for fBm conditioned to $X_0=0$, $X_{1/2}=1$ and $X_1=0$, for, from left to right, $H=0.1$, $H=0.25$ and $H=1$. The blue dashed-line represents the optimal paths when   neglecting the correlation between $[0,1/2]$ and $[1/2,1]$.}\label{OptimalPathFig}
\end{figure*}
\begin{equation}\label{ZmTOrder0}
Z_{0}^{\rm max}(m,T)=\frac{ m}{\sqrt{\pi}T^{\frac32}}e^{-\frac{m^2}{T}}\ ,
\end{equation}
which, after normalisation, allows to recover the distribution \eqref{MaxDistribBridge}.

The second order-$\varepsilon$ correction, which comes from the rescaling of the diffusive constant, is   obtained by replacing $T \to D_{\varepsilon,\tau}T$ in \Eqref{ZmTOrder0};  for the order-$\varepsilon$ term this gives
\begin{equation}
Z_{1B}^{\rm max}(m,T)=\frac{ z e^{-z^2}}{\sqrt{\pi}T} (2z^2-3)(1+\ln\tau)\ .
\end{equation}
Resumming these   corrections up to order $\varepsilon$ cancels all $\tau$ dependencies; recasting the relevant corrections into the power-law prefactor and the Gaussian tail and expressing the result in terms of the  dimensionless variable $y:=m/T^H$ finally yields
\begin{align}
{\cal P}^{\rm bridge}_{H}(m)&= 2\sqrt{\pi} T^H \big[Z_{0}^{\rm max}+\varepsilon (Z_{1A}^{\rm max}+Z_{1B}^{\rm max})\big]+\mathcal{O}(\varepsilon^2)\nn\\
&=\frac{2 y^{1-8\varepsilon}}{T^{H}}e^{- {y^2}{A_{\varepsilon}} + \varepsilon \mathcal{G}(y)+\mathrm{cst}}+\mathcal{O}(\varepsilon^2)\ .
\end{align}
The special function  $\mathcal{G}$ appearing here is as defined in Refs.~\cite{WieseMajumdarRosso2010,DelormeWiese2015,DelormeWiese2016},
\begin{align}
\mathcal{G}(y)=&-4 \left(y^2-1\right) \mathcal{J}\!\left(y^2\right)+2 \sqrt{\pi } e^{y^2} y\, \text{erfc}(y)\nn \\
&+2 y^2 \left[\log \left(4 y^2\right)+\gamma_{\rm E}\right]-4 \gamma_{\rm E} -2\ .
\end{align}
This result contains several non-trivial predictions: First, at small $m$, the distribution ${\cal P}_H^{\rm bridge}(m)$ has a power law   given by $ m^{1-8\varepsilon+\mathcal{O}(\varepsilon^2)}$.
This can be  obtained by considering the   probability starting at $m$ to remain positive (survive) up to time $T$, 
\be
{\cal S}(T,m) := \int_0^m \rmd m_1\,{\cal P}_{H}(m_1)\ .
\ee
In this relation the dependence of ${\cal P}_{H}(m)$ on $T$ is implicit. It is valid both for   the case of a bridge and  of  a free endpoint. 
To survive in a bridge in the limit of $m\to 0$ demands to survive both in the beginning and at the end, thus we expect that for small $m$
\be
{\cal S}^{\rm bridge}(T,m)  \sim \left[ {\cal S}^{\rm free}(T,m) \right]^2\ .
\ee
%
%
%
Using the result of Ref.~\cite{DelormeWiese2016} that $ {\cal P}_H^{\rm free}(m)\sim m^{\frac1H-2}$  implies that   
\begin{equation}
{\cal P}_H^{\rm bridge}(m)\sim m^{\frac2H -3}\ .
\end{equation}
This is in agreement with our order-$\epsilon $ result.

Second, at large $m$, ${\cal P}^{\rm bridge}_{H}(m)$ has a Gaussian tail with the   dimensionless variable $y^2=z^2/T^{2\varepsilon}=m^2/T^{2H}$ and a non-trivial number $A_{\varepsilon}= 1 + 4 \varepsilon  \log(2) + \mathcal{O}(\varepsilon^2)$. We will see in the next section why this number appears, and how we can compute it exactly (i.e.\ for all $H$).

Third, there is a crossover in the power-law behavior at large $y$, given by the asymptotic behaviour of the function ${\cal G}(y)$, 
\begin{equation}
\mathcal{G}(y) \underset{y \to \infty}{\simeq} 4 \ln(y)\ .
\end{equation}
This yields a subleading power-law behaviour at large $m$
\begin{equation}
{\cal P}_H^{\rm bridge}(m)\,  e^{A_{\varepsilon}\frac{m^2}{T^{2H}}} \sim m^{1-4\varepsilon+\mathcal{O}(\varepsilon^2)}\ .
\end{equation}

\subsection{Optimal path for fBm, and the tail of the maximum distribution}
In this section, we study the tail   of the maximum distribution for fBm. Contrary to a process with a free endpoint, the maximum is not taken at the end, and as a consequence the tail is not simply given by the known propagator evaluated at time $T$ at position $m$. 

We start with some general considerations:
If we choose $t_1, ..., t_n \in \mathbb{R}$, then the density distribution for a fBm path $X_t$ to take values $X_{t_1}=x_1, ...,  X_{t_n}=x_n$ can be expressed,  using the Gaussian nature of the process $X_t$, as
\begin{equation}\label{GaussianProba}
{\cal P}_n(x_1,x_2,...,x_n) = \exp \!\left(- \frac12 \sum\limits_{ij} x_i \mathcal{M}_{ij} x_j \right)\ .
\end{equation}
The matrice $\mathcal{M}_{ij}$ is given by
\begin{equation}\label{GaussianMatrix}
\mathcal{M}^{-1}_{ij} = \langle X_{t_i} X_{t_j}\rangle = t_i^{2H}+ t_j^{2H}- |t_i -t_j|^{2H}\ .
\end{equation}
To study bridges, consider now two points, $x_1 = x$ at time 
$t_1=t$ with $0<t<T$ and $x_2 = 0$ at time $t_2=T$. The probability distribution of $x$ given $x_T=0$ is then given by
\begin{equation}\label{104}
{\cal P}(x_{t}= x |x_T=0)= {\cal P}_2(x,0) = \exp \left(  -\frac { {\cal M}_{11}\, x^2}2 \right) \ .
\end{equation}
The matrix element in question is (with $\vartheta = t/T$)
\be\label{105}
\frac{\mathcal{M}_{11}}{2} =  \frac1{T^{2H}}  \frac{1}{4 \vartheta ^{2 H}-\left[\vartheta ^{2
   H}-(1-\vartheta )^{2 H}+1\right]^2}
\ .
\ee
It takes its minimum for $\vartheta=\frac 12$. The tail for the maximum of a bridge is thus given by \Eqref{104} with the matrix element ${\cal M}_{11}$ in Eq.~(\ref{105}) evaluated at $\vartheta=\frac12$: 
%
%
%
%
\begin{align}
{\cal P}_T(m)&\approx {\cal P}(x_{T/2}=m|x_T=0)\nn\\
&=e^{-\frac{m^2}{T^{2H}}\frac{4^H}{4-4^H} +\mathcal{O}(\log(m))}\ .
\end{align}
This heuristic argument is consistent with the result from our $\varepsilon$ expansion, and allows us to predict the {\em exact} value of the constant $A_{\varepsilon}$,
\begin{equation}
A_{\varepsilon}=\frac{4^H}{4-4^H} = 1 + 4 \ln(2) \varepsilon + \mathcal{O}(\varepsilon^2)\ .
\end{equation}
We can go further and study the shape of the optimal path with conditions $X_0=X_1=0$ and $X_{1/2}=1$. This is done by considering ${\cal P}_n(x,1,0)$, taken at time $t_1=t$, $t_2=1/2$ and $t_3=T=1$. We then find $X^{\rm SP}_t=x$ which minimises the ``energy'' $-\ln {\cal P}_3(x,m,0)$. This is for $0\le \vartheta\le \frac12$ achieved for 
\be
X_t^{\rm SP}\!\! = \frac{m}{4{-}4^H}\Big[2 -2 (1-2\vartheta )^{2 H}+4^H
   (1-\vartheta )^{2
   H}+4^H
  \vartheta ^{2
   H}-4^H \Big].
\ee
For $\frac T 2 < t\le T$ one has $X_t^{\rm SP}= X_{T-t}^{\rm SP}$. 
This is represented for $m=1$ and $T=1$ in red in Fig.~\ref{OptimalPathFig} for various values of $H$.
It is interesting to observe that this {\em optimal path} is not a straight line going from $X_0=0$ to $X_{1/2}=1$ and back to $X_1=1$, but  at $t=1/2$ peaked for $H<1/2$, and smoothened for $H>1/2$. 
It is equivalently interesting to compare this to the optimal path which goes from $X_0=0$ to $X_{1/2}=1$, without imposing any constraint at $t=1$, plus a similar segment from $X_{1/2}=1$ to $X_{1}=0$ without constraint on $X_0$ (blue dashed lines). This would indeed be the optimal path  if there were no correlations between times $t<1/2$ and $t>1/2$.

We finally note that the limit of $H\to 1$ is non-trivial, and given by (see right of Fig.~\ref{OptimalPathFig})
\bea
X_t^{\rm SP} &=& \frac m{{\log (4)}} \Big\{ (1-2 \vartheta )^2 \log (1-2 \vartheta)-2
   (1-\vartheta)^2 \log (1-\vartheta)\nn\\
   && +\vartheta [\log
   (16) -2 \vartheta
   \log (4 \vartheta)] \Big\} \ , \qquad 0\le t\le \frac T2
\eea
and $X_t^{\rm SP}=X_{T-t}^{\rm SP}$ for $\frac T 2<t<T$.
We expect this also to be  the lowest-energy fluctuation for the fBm bridge.

\begin{figure*}[t]
\centerline{\!\fig{5.9cm}{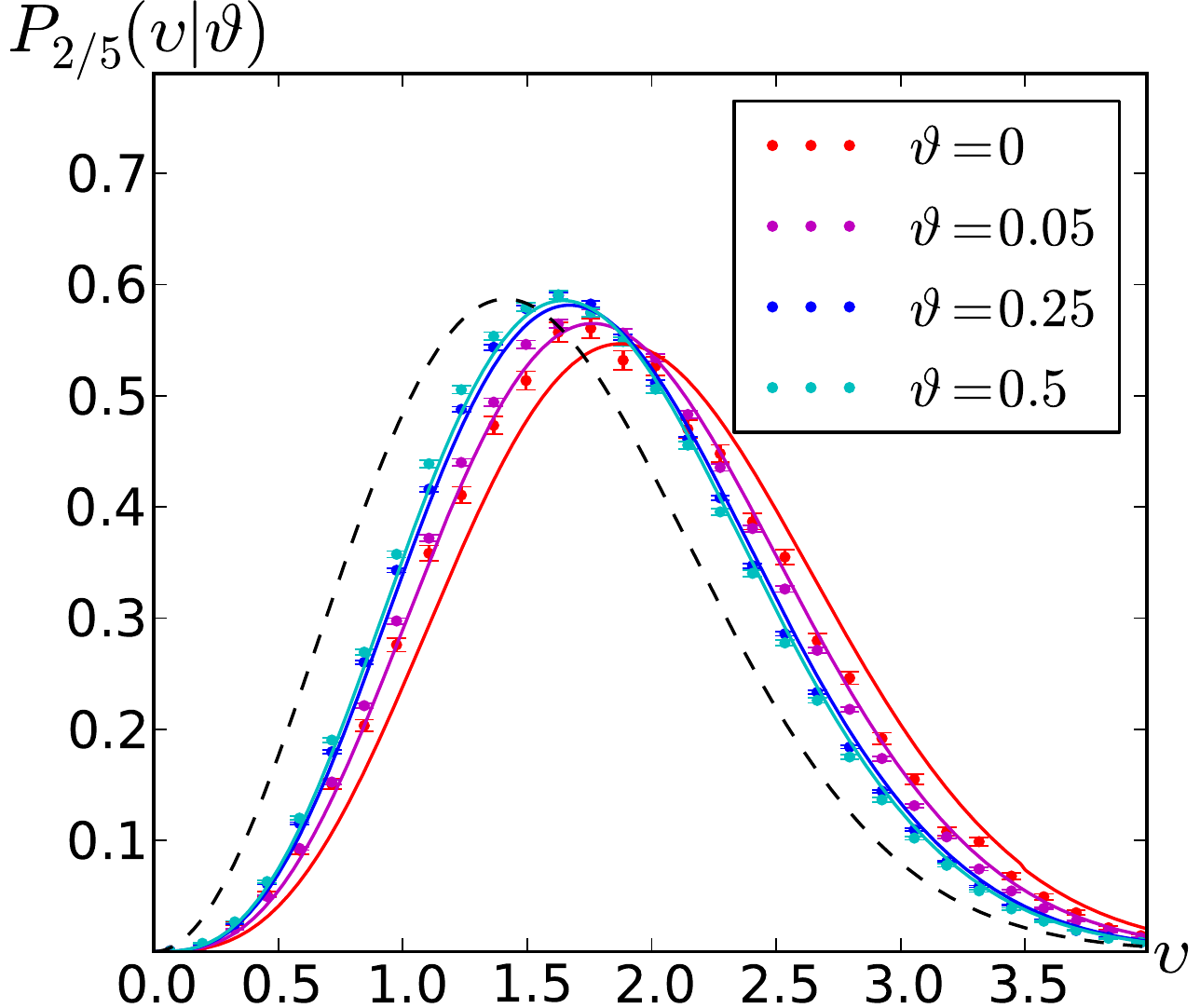}
        \fig{5.9cm}{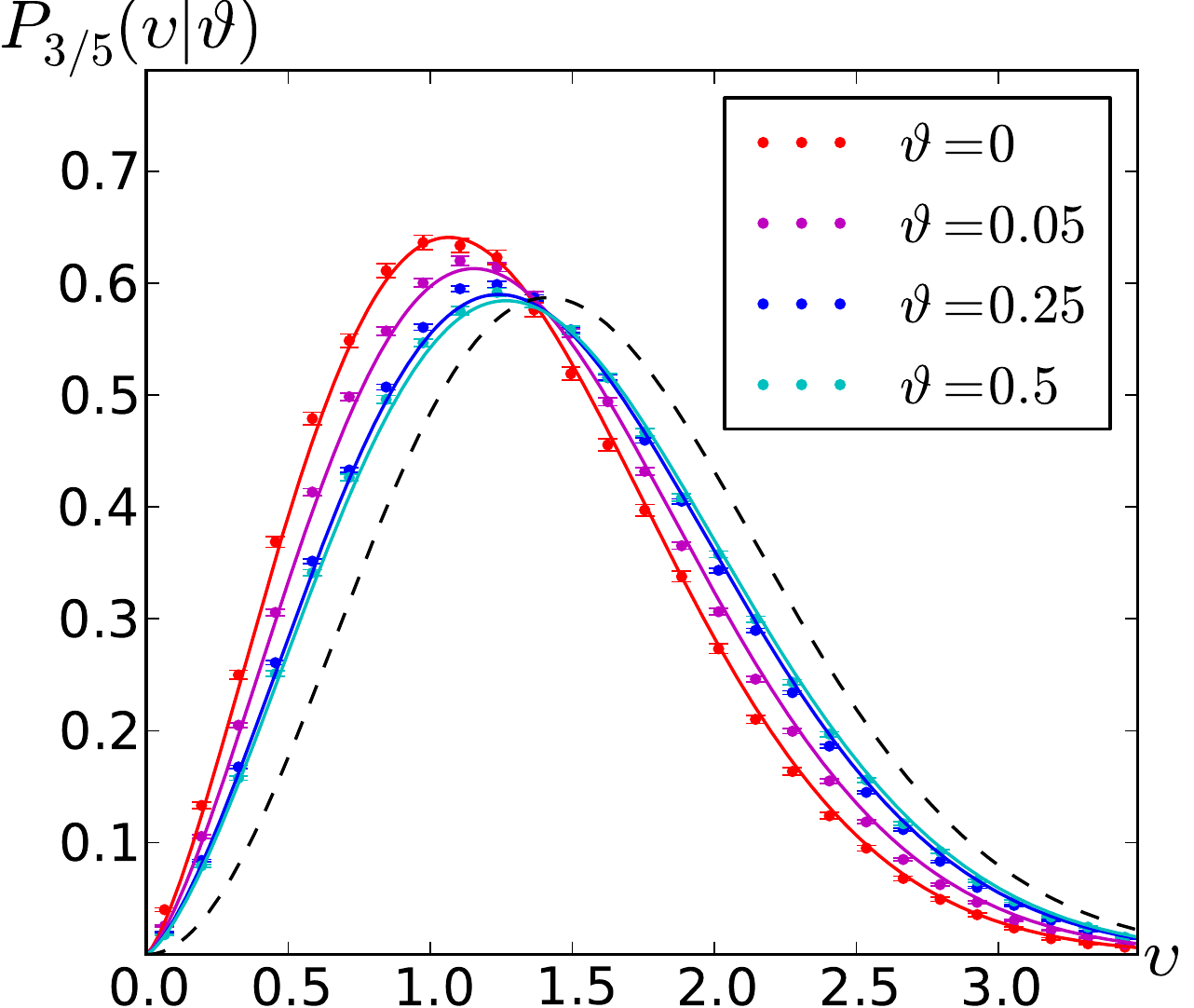} 
        \fig{5.9cm}{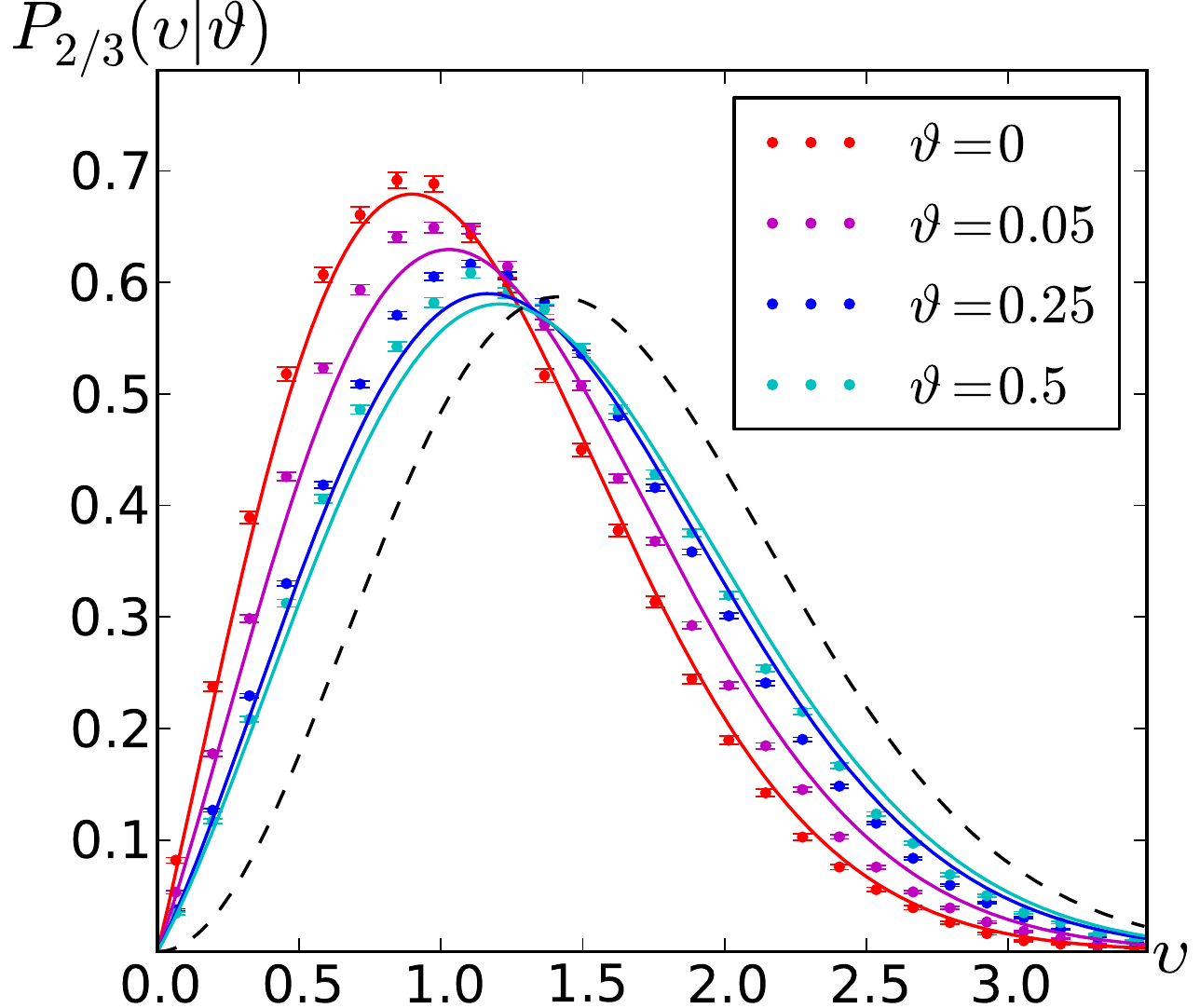}} 
        \caption{Numerical results for $P_H(\upsilon| \vartheta)$ for $H=\frac25$ (left), $H=\frac35$ (middle) and $H=\frac23$ (right). The values of $ \vartheta$ are chosen as $ \vartheta=0$, $ \vartheta=0.05$, $ \vartheta=0.25$ to $ \vartheta=0.5$, the maximum useful value due to the symmetry $\vartheta \to 1-\vartheta$.    We used $N=2^{18}$ points, and $5 \times 10^6$ samples.}
        \label{f:conditional}
\end{figure*}

\subsection{Joint Distribution of $m$ and $t_{\rm max}$}
To obtain the joint distribution of $m$ and $t_{\rm max}$,  we start with Eq.~(44) of Ref.~\cite{DelormeWiese2016}, and specify $m_1=m_2=m$. This is equivalent, in the notations of \cite{DelormeWiese2016}, to setting
\begin{equation}
y_1=\frac{m}{\sqrt{2} \vartheta^H}\ ,\ y_2=\frac{m}{\sqrt{2} (1-\vartheta)^H}\ \text{ where } \vartheta=\frac{t_{\rm max}}{T}\ .
\end{equation}
The resulting expression can more compactly  be written in terms of  
\be \label{y}
{\upsilon}:= \frac{m }{\sqrt{2} [\vartheta(1-\vartheta)]^{H}}\ .
\ee
Recasting terms proportional to   $\ln(\vartheta)$, $\ln(1-\vartheta)$ and $\ln(\upsilon)$  into the prefactor, we get 
\begin{align}\label{ZplusJoint}
&       Z^+(m,\vartheta ;x_0;m,1-\vartheta)\nn\\
& =\frac{x_0^{2-4\varepsilon} \upsilon^{2-8\varepsilon} e^{-\frac{\upsilon^2}{2}}}{2 \pi [\vartheta(1-\vartheta)]^{3H-1}}\Big\{ 1+\varepsilon\Big[ {\cal F}(\upsilon,\vartheta)+C_2 \big] \Big\}+\mathcal{O}(\varepsilon^2)
        \end{align}
with
\begin{align}\label{Fjoint}
{\cal F}(\upsilon,\vartheta) =\;&    \frac{\mathcal{I}\!\big(\upsilon(1-\vartheta)\big)+\mathcal{I}\!\big(\upsilon \vartheta\big)-\mathcal{I}(\upsilon)+2(\upsilon^2-1)}{\upsilon^2 \vartheta (1-\vartheta)} \nn\\
& -\frac{\mathcal{I}\!\big(\upsilon(1-\vartheta)\big)}{1-\vartheta} -\frac{\mathcal{I}\!\big(\upsilon \vartheta\big)}{\vartheta} +2\,\mathcal{I}\!\left(\upsilon \sqrt{1-\vartheta}\right) \nn\\
& +2\,\mathcal{I}\!\left(\upsilon \sqrt{\vartheta}\right) + \upsilon^2 \big(\ln(2\upsilon^2) + \gamma_{\rm E}\big)\nn\\
& -12-8\ln(2)   \ , \\
C_2 =\;&  4 \Big[ 2-\gamma_{\rm E} +\log (2)\Big]
\end{align}
First, this result allows us to recover Eqs.~\eqref{CorrectionTmaxDistrib} and \eqref{TmaxScalingF}, noting that 
\be
{\cal F}^{\rm max}(\vartheta) = \sqrt{\frac2\pi}\int_0^\infty \rmd \upsilon\,  {\upsilon^2} {e^{-\frac{\upsilon^2}2}}{\cal F}(\upsilon,\vartheta)\ .
\ee
As we   defined $\int_0^1\rmd \vartheta \, {\cal F}^{\rm max}(\vartheta) =0$, there is an additional constant $C_2$, related to the prefactor $\upsilon^{-8\varepsilon}$ in \Eqref{ZplusJoint}.

Second, we can extract the conditional probability  of $\upsilon$,  given $\vartheta$. This is interesting  since for a   Brownian the latter depends  only on the variable $\upsilon$ introduced in Eq.~(\ref{y}),
\begin{equation}\label{ConditionnedBrownian}
{\cal P}^{\rm bridge}_{H=\frac12}(\upsilon |\vartheta)= \sqrt{\frac{2}{\pi}} \upsilon^2 e^{-\frac{\upsilon^2}{2}}\ .
\end{equation}
For a generic value of $H=\frac12 + \varepsilon$, our $\varepsilon$ expansion, recast in an exponential form, gives
\begin{eqnarray}\label{ConditionnedResult}
{\cal P}^{\rm bridge}_H(\upsilon| \vartheta)&=& \sqrt{\frac{2 }{\pi}} \upsilon^{\frac{2}{H}-2} e^{-\frac{\upsilon^2}{2}+\varepsilon [ \mathcal{F}(\upsilon,\vartheta) +C_2-  \mathcal{F}^{\rm max}( \vartheta) ]}\nn\\
&& +\mathcal{O}(\varepsilon^2)\ .
\end{eqnarray}
The functions $\mathcal{F}( \upsilon,\vartheta) $ and $\mathcal{F}^{\rm max}( \vartheta) $ are defined in Eqs.~(\ref{Fjoint}) and (\ref{TmaxScalingF}).
The exponent in \Eqref{ConditionnedResult} can be derived from scaling. 
To this aim, note that  the probability to have a maximum of $m$ up to time $T$ is
\be
{\cal P}_H(m)   = \partial_{m }{\cal S}(T,m)  \ .
\ee
On the other hand, the probability that the maximum $m$ is taken at   time $T$ is
\be
{\cal P}_H(m|T)  = \partial_{T} {\cal S}(T,m)\ .
\ee 
We conclude that for small $m$
\be
{\cal P}_H^{\rm bridge} (m|T) \sim \frac mT {\cal P}^{\rm bridge}_{H}(m) \sim   m^{\frac2H -2} \sim \upsilon^{\frac2H -2} \ .
\ee
This exponent, written in Eq.~(\ref{ConditionnedResult}), agrees with the perturbative expansion
\begin{equation}
\frac{2}{H}-2 = 2 - 8 \varepsilon + \mathcal{O}(\varepsilon^2)\ .
\end{equation}
Finally, using the result \eqref{105}, and expressing it in terms of $\upsilon$ predicts a tail $e^{- A'_\epsilon \upsilon^2 }$, with 
\bea\label{Aprime}
 A'_\epsilon &=&  \frac{2 [\vartheta(1-\vartheta)]^{2H} }{4 \vartheta ^{2 H}-\left[\vartheta ^{2
   H}-(1-\vartheta )^{2 H}+1\right]^2} \\
   & = & \frac12\left[ 1+ \epsilon^2  \frac{[ ( 1-\vartheta ) \log (1- \vartheta )+ \vartheta  \log
   ( \vartheta )]^2}{2 (1- \vartheta )  \vartheta } + {\cal O}(\epsilon^3)\right] \nn
\ .
\eea
Thus our resummation (\ref{ConditionnedResult}) is correct to order $\epsilon$; whether at higher order it is preferential to use $\upsilon$ introduced in \Eqref{y} with $A_\epsilon'$ given in \Eqref{Aprime}, or whether one should keep $e^{-\upsilon^2/2}$ for the tail and redefine $\upsilon$ can only be answered after a second-order calculation.

We verified
the prediction (\ref{ConditionnedResult}) for ${\cal P}^{\rm bridge}_H(\upsilon| \vartheta)$  numerically, see Fig.~\ref{f:conditional}. The agreement is   good for $H$ close to $\frac12$, both for $ \epsilon = -\frac1{10}$    and $ \epsilon = \frac1{10}$ (left two figures). Corrections of order $\epsilon^2$ can be anticipated, since our numerical results for both  $\epsilon = - \frac1{10}$  and $\epsilon =   \frac1{10}$  show approximately the same (small) deviation from the analytics, independent of the  sign of $\epsilon$.

These putative $ \mathcal{O}(\epsilon^2)$ corrections   also explain the larger  systematic deviations  for $H=\frac23$, i.e.\ $\epsilon=\frac16$ (right plot).

\section{Conclusions}

In this article we developed a systematic analytical  framework to treat bridge processes for fractional Brownian motion, in an expansion around Brownian motion. We considered the probability of the time $t_+$ that a bridge process is positive, and of the    time $t_{\rm max}$  it achieves its maximum. For a Brownian bridge, both  $t_+$ and $t_{\rm max}$ have the same uniform probability distribution. For  a fractional Brownian bridge, both observables have the same power-law behavior for times close to the beginning and   end, but the subleading scaling functions are rather different. We calculate them to first order in $\epsilon$, and verified them to high precision with numerical simulations. We also obtained and checked  the joint distribution of the maximum $m$, and the time $t_{\rm max}$ when this maximum is taken.  These tests were possible due to the development of an efficient algorithm to generate samples of fBm bridges. 

\section{Acknowledgments}
We thank P.\ Krapivsky, K.\ Mallick, A. Rosso and
T. Sadhu for stimulating discussions, and PSL for support
through grant ANR-10-IDEX-0001-02-PSL.
\appendix

\section{Details on correlation functions for the bridge}
\label{a:bridge-details}
Starting from Eqs.~(\ref{Bridge_def1}) and (\ref{Bridge_def}), and inserting the identity $ \delta(x)  = \int_{-\infty}^{\infty} e^{i k x} \frac{\rmd k}{2\pi}$, we obtain
\begin{align}
\langle    \delta(X_T-a)\rangle &= \int_{-\infty}^{\infty} { \frac{\rmd k}{2\pi}}\,\left<     e^{i k (X_T-a)}\right>\nn \\
&=\int_{-\infty}^{\infty} { \frac{\rmd k}{2\pi}}\,e^{ - i k a}\   e^{- \frac{k^2}{2} \langle X_T^2 \rangle }  \nn \\
&={  \frac{e^{-\frac{a^2}{2 \langle X_{T}^2 \rangle}}}{\sqrt{2\pi}\sqrt{\langle X_{T}^2 \rangle}}} \ ,
\end{align}
\begin{align}
\langle X_{t_1}   \delta(X_T-a)\rangle &= \int_{-\infty}^{\infty} { \frac{\rmd k}{2\pi}}\,\left< X_{t_1}   e^{i k (X_T-a)}\right>\nn \\
&=\int_{-\infty}^{\infty} { \frac{\rmd k}{2\pi}}\,e^{ - i k a}\, i k\, \langle X_{t_1} X_{T} \rangle    e^{- \frac{k^2}{2} \langle X_T^2 \rangle }  \nn \\
&={  \frac{e^{-\frac{a^2}{2 \langle X_{T}^2 \rangle}}}{\sqrt{2\pi}\sqrt{\langle X_{T}^2 \rangle}}} 
\frac{a \left<X_{t_1}X_T \right>}{ \left<X_T^2 \right> }\ ,
\end{align}
\begin{align}
 \langle  X_{t_1}& X_{t_2} \delta(X_T-a)\rangle = \int_{-\infty}^{\infty} { \frac{\rmd k}{2\pi}}\,\left< X_{t_1} X_{t_2} e^{i k (X_T-a)}\right>\nn \\
=&\int_{-\infty}^{\infty} { \frac{\rmd k}{2\pi}}\,e^{ - i k a}  e^{- \frac{k^2}{2} \langle X_T^2 \rangle }   \nn \\
& \times \Big [\langle X_{t_1} X_{t_2} \rangle  -k^2 \langle X_{t_1} X_{T} \rangle \langle X_{t_2} X_{T} \rangle  \Big] \\
=&{  \frac{e^{-\frac{a^2}{2 \langle X_{T}^2 \rangle}}}{\sqrt{2\pi}\sqrt{\langle X_{T}^2 \rangle}}} \nn \\
& \times \left[ \langle X_{t_1} X_{t_2}\rangle  +{  \Big(  a^2 -\langle X_{T}^2\rangle  \Big) }\frac{\langle X_{t_1} X_{T} \rangle\langle X_{t_2} X_{T} \rangle}{\langle X_{T}^2\rangle^2}\right]\ .\nn
\end{align}
From the first to the second line of the last two equations we used Wick's   theorem and the fact that $X_t$ has  mean zero. Putting everything together, we arrive at Eqs.~(\ref{10}) and (\ref{11}).

\section{Abel transform and inversion of $\mathcal{K}_{\frac12}$ transform} \label{a:Abel}
For a real function $g(\vartheta)$ non-vanishing on the interval $[0,1]$, we consider the transformation $\mathcal{K}_{\frac12}$ defined as
\begin{equation}\label{transfoDef}
\bar g(\kappa)\equiv \mathcal{K}_{\frac12}[g](\kappa):= \int_0^1\! \frac{g(\vartheta)}{\sqrt{1+\kappa \vartheta}}\,\rmd \vartheta\ .
\end{equation}
The question is how to reconstruct   $g$, knowing  $\bar g$.

The Abel transform  $F$ of a function $f$ is defined as \cite{Abel1826,BracewellBook}
\begin{equation}\label{A2}
F(y)= \int_y^{\infty}\!\!\frac{2 r f(r)}{\sqrt{r^2-y^2}}\,\rmd r\ .
\end{equation}
The inverse formula, allowing to recover $f$ from $F$, is
\begin{equation}\label{Abel-inverse}
f(r)=-\frac{1}{\pi}\int_r^{\infty}\!\!\frac{F'(y)}{\sqrt{y^2-r^2}}\,\rmd y\ .
\end{equation}
To make the link with $\mathcal{K}_{\frac12}$, we change variables from $\vartheta$ to $r:= \sqrt{\vartheta}$ in \Eqref{transfoDef}, and introduce $f(r):= g(\vartheta = r^2)$. Then, for $\kappa>0$, 
\begin{equation}
\bar{g}(\kappa) = \int_0^1 \!\frac{ f(r)}{\sqrt{1+ \kappa r^2}}\,2 r \,\rmd r = \frac{2}{\sqrt{\kappa}}\int_0^{\infty}\!\!\frac{ f(r) r}{\sqrt{\frac{1}{\kappa}+r^2}}  \,\rmd r\ .
\end{equation}
In the last equality, we changed the upper integration limit, using $f(r)=0$ for $r>1$. 
We now continue $\bar{g}(\kappa) \sqrt{\kappa}$ in the complex plane from real positive to real negative $\kappa$, by setting $\kappa = e^{i \varphi}/y^2|_{\varphi=\pm \pi}$ with $y>0$. 
This  gives
\begin{equation}
\begin{split}
\bar{g}(\kappa) \sqrt{\kappa} =&\,\int_y^{\infty}\!\!\frac{2 r f(r)}{\sqrt{r^2-y^2}}\,\rmd r + \int_0^y\!\frac{2 r f(r)}{\sqrt{r^2-y^2}}\,\rmd r\\
=&\,F(y)+ e^{-i \varphi/2} G(y)\ .
\end{split}
\end{equation}
We have split the integral over $r$ into two parts: the first part is a real function  $F(y) \in \mathbb{R}$, which is   the Abel transform of $f(r)$.  The second term is purely imaginary because of the denominator; which of the two possible branches is taken depends on how we continued $\bar g(\kappa)\sqrt{\kappa}$, choosing either of  the branches $\varphi =\pm \pi$. This means that we can express the Abel transform $F(y)$ of $f(r)$ from $\bar g(\kappa)$ as
\begin{equation}\label{A6}
F(y)= \mathfrak{R}\!\left[\left.\bar g(\kappa) \sqrt{\kappa} \right|_{\kappa=-1/y^2}\right]\ ,
\end{equation}
where $\mathfrak{R}$ denotes the real part. 
We can now use       formula (\ref{Abel-inverse}) to invert the  Abel transform.

Since $f(r)$ vanishes for $r>1$, according to the definition (\ref{A2}) also $F(y)$ vanishes for $y>1$. One can thus reduce the upper bound in \Eqref{Abel-inverse} to 1. 
Finally reintroducing the function $g(\vartheta)$ instead of $f(r)$, we  get
\begin{equation}\label{A8}
g(\vartheta)=-\frac{1}{\pi}\int_{\sqrt{\vartheta}}^{1} \frac{F'(y)}{\sqrt{y^2-\vartheta}} \,\rmd y
\ ,
\end{equation}
where $F(y)$ is defined from $\bar{g}(\kappa)$ in \Eqref{A6}. We now want to apply this to compute $g_1(\vartheta)$ from Eq.~\eqref{g1Inverse}. We need to compute the inverse $\mathcal{K}_{1/2}$ transform  of
\begin{align}
\bar g^{\rm pos}_1(\kappa)&-\bar g_{0,1}(\kappa)=  8 \left(\frac{1}{\sqrt{\kappa+1}}+1\right) \log \left(\sqrt{\kappa+1}+1\right) \nn \\
&-16\frac{ \log (4)-1}{\sqrt{\kappa+1}+1}-\frac{4 \left(\kappa+\sqrt{\kappa+1}+1\right) \log (\kappa+1)}{\kappa}.
\end{align}
From scaling, we expect that close to the boundary
\be
g_1(\vartheta) \simeq - 2 \ln\Big(\vartheta (1-\vartheta)\Big)\ .
\ee  To simplify the calculation, we subtract this divergent part. Define\checked
\begin{align}
\bar g^{\log}(\kappa):=& \int_0^1 \rmd \vartheta\, \frac{\ln\big(\vartheta(1-\vartheta)\big)+2}{\sqrt{1+\kappa \vartheta}}\nn\\
=&\,\frac{4  [\log (2)-1]}{ 1+\sqrt{\kappa +1}  } +\frac{2 \sqrt{\kappa +1} \log (\kappa +1)}{\kappa }\nn\\&
+\frac{4\left(1- \sqrt{\kappa +1}\right) \log \left(\sqrt{\kappa +1}+1\right)}{\kappa } \ .
\end{align}
Setting $\bar g(\kappa):= \bar g_1^{\rm pos}(\kappa)-\bar g_{0,1}(\kappa)+2 \bar g^{\ln}(\kappa)$ in \Eqref{A6} yields 
\begin{align}
F(y) = &-\frac{8 y^2 \log (y)}{\sqrt{1-y^2}}-24 \sqrt{1-y^2} \log (2)\nn\\
&-\frac{8 \left(y^2-1\right) \arcsin(y)}{y} \ .
\end{align}
Computing the integral \eqref{A8} finally gives  \checked
\begin{align} \nn
g(\vartheta) & = \mathcal{K}_{\frac12}^{-1}\left[\bar g_1(\kappa)-\bar g_{0,1}(\kappa)+2 \bar g^{\ln}(\kappa)\right]\\
&=4\Bigg[3-\frac{1}{\sqrt{1-\vartheta}+1} -\frac{1}{\sqrt{\vartheta}+1} \nn \\
&\;\;\;\;\;\;\;+\ln\left(\frac{(\sqrt{\vartheta}+1)(\sqrt{1-\vartheta}+1)}{16}\right) \Bigg]\ . 
\end{align}
Adding the logarithmic terms, we recover the result \eqref{ResultPosTimeOrdre1} given in the main text.
 
\section{Inverse Laplace transforms necessary for the maximum of the bridge, and other useful relations}
\label{a:inv-Lap-trafos}\nopagebreak
In this appendix we give a table of   useful relations for the inverse Laplace transforms encountered in this article. 

All appearing hypergeometric functions can be eliminated by using two special functions, introduced in Refs.~\cite{WieseMajumdarRosso2010,DelormeWiese2015,DelormeWiese2016}, and named ${\cal I}(x)$ and ${\cal J}(x)$, 
\begin{align}
\mathcal{I}(x) &=  \frac{1}{6} x^4 \, _2F_2\left(1,1;\frac{5}{2},3;\frac{x^2}{2}\right)+\pi 
   \left(1-x^2\right) \text{erfi}\!\left(\frac{x}{\sqrt{2}}\right)\nn\\ 
   &+\sqrt{2 \pi }
   e^{\frac{x^2}{2}} x+2-3
x^2\ ,\\
\mathcal{J}(x) &= \frac{1}{2} \pi\,  \text{erfi}\left(\sqrt{x}\right)-x \, _2F_2\!\left(1,1;\frac{3}{2},2;x\right)
\end{align}
These functions are related to each other by the relations 
\begin{align}
{\cal I}(x) &= 2+2(1-x^2)\, {\cal J}\!\left( \frac{x^2}{2}\right) + \sqrt{2\pi } e^{\frac{x^2}2}x \
\mbox{erfc}\!\left(\frac{x}{\sqrt 2}\right)\ , \\
  {\cal I}(x) &= -2\, e^{\frac{x^2}2}\partial_x^2\left[ e^{-\frac{x^2}2}  {\cal J}\!\left(\frac{x^2}{2}\right) \right]\ .
\end{align}
To arrive at these identities, and to express everything in terms of one of these two functions, two non-trivial relations between hypergeometric functions were used (they can be checked by Taylor-expansion to high order)
\begin{align}
&-3 \, _2F_2\left(1,1;\frac{3}{2},2;\frac{x^2}{2}\right)+\,
_2F_2\left(1,1;2,\frac{5}{2};\frac{x^2}{2}\right)\nn\\ &+\frac 6{x^2}
\left[\sqrt{\frac\pi {2} }\, \frac{e^{\frac{x^2}{2} }}x \,
\text{erf}\left(\frac{x}{ \sqrt{2}}\right)-1\right]=0
\end{align}\begin{align}
&-x^3 \left[ 3 \,
_2F_2\left(1,1;\frac{3}{2},2;-\frac{x^2}{2}\right)+\,
_2F_2\left(1,1;2,\frac{5}{2};\frac{x^2}{2}\right)\right]\nn\\
&+\text{erf}\left(\frac{x}{\sqrt
        {2}}\right) \left[3 \pi  x
\text{erfi}\left(\frac{x}{\sqrt{2}}\right)-3
\sqrt{2 \pi } e^{\frac{x^2}{2}}\right]+6 x = 0\ .
\end{align}
We now express the needed inverse Laplace transforms     either in terms of $\cal I$ or $\cal J$, depending on which form is more compact. (Note that each function appears naturally in a given context \cite{DelormeWiese2016}). 

Transforms  involving only $e^{-\sqrt{s}}$, and powers of $\sqrt s$ are elementary,  
\bea
\mathcal{L}^{-1}_{{s \to t}} \Big[ e^{-\sqrt{s}}\Big] &=& \frac{e^{-\frac{1}{4 t}}}{2 \sqrt{\pi } t^{3/2}} \\
\mathcal{L}^{-1}_{{s \to t}} \Big[ e^{-\sqrt{s}} \sqrt{s}\Big] &=& -\frac{e^{-\frac{1}{4 t}} (2 t-1)}{4 \sqrt{\pi } t^{5/2}} \\
\mathcal{L}^{-1}_{{s \to t}} \Big[\frac{e^{-\sqrt{s}}}{\sqrt{s}} \Big] &=& \frac{e^{-\frac{1}{4 t}}}{\sqrt{\pi t }}  
\ .\eea
\begin{widetext}\noindent
Transforms with an additional factor of $\ln (s)$ are
\bea
\mathcal{L}^{-1}_{{s \to t}} \Big[e^{-\sqrt{s}} \sqrt{s} \log (s) \Big] &=&  -\frac{e^{-\frac{1}{4 t}}}{4 \sqrt{\pi } t^{5/2}}  \left\{  -2 t\;  {\cal I}\!\left(\frac{1}{\sqrt{2
t}
    }\right) +(2 t-1) \Big[\log (4 t)+\gamma_{\rm E} \Big]  \right\}
\\
\mathcal{L}^{-1}_{{s \to t}} \Big[\frac{e^{-\sqrt{s}} \log (s)}{\sqrt{s}} \Big] &=&\frac{ e^{-\frac{1}{4 t}}}{\sqrt{\pi t }  } \left[ 2\,{\cal J}\!\left(\frac{1}{4t} \right) -\log (4 t)-\gamma_{\rm E}\right] \\
\mathcal{L}^{-1}_{{s \to t}} \Big[ e^{-\sqrt{s}} \log (s) \Big] &=& \frac{ e^{-\frac{1}{4 t}}}{4 \sqrt{\pi } 
        t^{5/2}}  \left[ 2\,{\cal
J}\!\left(\frac{1}{4t} \right) -\log (4 t)-\gamma_{\rm E}\right]-\frac{\text{erfc}\left(\frac{1}{2 \sqrt{t}}\right)}{t}
\ .\eea
Transforms involving the
exponential integral function are   
\bea
\mathcal{L}^{-1}_{{s \to t}} \Big[ \text{Ei}\left(-\sqrt{s }\right)
\Big]&=&  -\frac{\text{erfc}\left(\frac{1}{2
                \sqrt{t}}\right)}{2 t} \\
\mathcal{L}^{-1}_{{s \to t}} \Big[ e^{\sqrt{s }} \text{Ei}\left(-2
\sqrt{s }\right)\Big] &= & \frac{e^{ -\frac{1}{4t}}}{4 \sqrt{\pi
} t^{3/2}}  { \left[ 2\, {\cal J} (\frac{1}{4 t} )+\log
   (t)-\gamma_{\rm E} \right]}{}-\frac{\text{erfc}\left(\frac{1}{2
   \sqrt{t}}\right)}{2 t} \\
\mathcal{L}^{-1}_{{s \to t}} \Big[ \sqrt s e^{\sqrt{s }} \text{Ei}\left(-2
\sqrt{s }\right)\Big] &=& \frac{e^{ -\frac{1}{4 t} }}{8 \sqrt{\pi
}
   t^{5/2}}  { \left\{ 2 t\;  {\cal I}\!\left(\frac{1}{\sqrt{2
t}
    }\right) +(2t-1)\Big[\ln(t)-\gamma_{\rm E}\Big]-8t\right\}}
\\ \mathcal{L}^{-1}_{{s \to t}} \Big[\frac{e^{\sqrt{s}} \text{Ei}\left(-2
\sqrt{s}\right)}{\sqrt{s}}\Big] &=& \frac{e^{ -\frac{1}{4 t}
}}{2 \sqrt{\pi t
}}  { \left[ \gamma_{\rm E}- 2 \;  {\cal J}\!\left(\frac{1}{{4
t}
    }\right)- \ln(t)\right]}
\ .\eea

\end{widetext}

\tableofcontents
%

\end{document}